\documentclass[english]{article}
\usepackage[T1]{fontenc}
\usepackage[latin9]{inputenc}
\usepackage{geometry}
\usepackage{verbatim}
\usepackage{amsmath}
\usepackage{amssymb}
\usepackage{graphicx}
\usepackage[authoryear]{natbib}

\makeatletter

\usepackage{mystyle}

\@ifundefined{showcaptionsetup}{}{%
 \PassOptionsToPackage{caption=false}{subfig}}
\usepackage{subfig}
\makeatother

\usepackage{babel}
\begin{document}

\title{An Information-Theoretic Test for Dependence \\
with an Application to the Temporal Structure\\
of Stock Returns}

\author{Galen Sher \\
Department of Economics \\
Oxford-Man Institute \\
University of Oxford \and 
Pedro Vitória%
\thanks{Corresponding author. Email: $\mathtt{pedro.vitoria@maths.ox.ac.uk}$. %
}~%
\thanks{The authors would like to acknowledge research support from the Oxford--Man Institute of Quantitative Finance. The second author gratefully acknowledges the support of FCT doctoral grant
SFRH / BD / 68331 / 2010.%
}\\
Mathematical Institute \\
Oxford-Man Institute\\
University of Oxford}
\maketitle
\begin{abstract}
Information theory provides ideas for conceptualising information
and measuring relationships between objects. It has found wide application
in the sciences, but economics and finance have made surprisingly
little use of it. We show that time series data can usefully be studied
as \emph{information }-- by noting the relationship between statistical
redundancy and dependence, we are able to use the results of information
theory to construct a test for joint dependence of random variables.
The test is in the same spirit of those developed by \cite{ryabko2005application, ryabko2006universal, ryabko2006universal2}, but differs from these in that we add extra randomness to the original stochatic process.
It uses data compression to estimate the entropy rate of a stochastic process, which
allows it to measure dependence among sets of random variables, as
opposed to the existing econometric literature that uses entropy and
finds itself restricted to pairwise tests of dependence. 
We show how serial dependence may be detected in S\&P500 and PSI20
stock returns over different sample periods and frequencies. We apply
the test to synthetic data to judge its ability to recover known temporal
dependence structures.

\bigskip{}
{\emph{Keywords}:} Information theory, entropy, data compression,  independence test
\end{abstract}

\section{Introduction}

We apply some concepts from information theory to the study of time
series economic phenomena. Asset pricing and statistical models of
stock returns try to capture the salient features of the latent processes
that generate our observed data. Information theory, following \citet{Shannon1948},
provides ideas for conceptualising information and measuring relationships
between collections of objects that are highly relevant for describing
the behaviour of economic time series. In finance, a time series of
stock returns is a collection of real numbers that are ordered in
time. The real numbers have a natural ordering, 
and using ideas from information theory to study the relationship
between this ordering of returns and the time ordering of returns
brings out salient features of the latent generating processes to
be modelled.

In particular, we may ask whether there are information redundancies
over time in price changes in general and stock returns in particular.
\citet{Hayek1945} argued that prices in an economy are a mechanism
for communicating information that helps people to organise their
behaviour, so the informativeness or uninformativeness of prices has
important implications for the function or dysfunction of a capitalist
system. The efficient market hypothesis (EMH) of \citet{Fama1965}
argues that asset prices reflect all available information. Since
asset prices are merely summary measures of an enormous variety of
market conditions, and since acting on information incurs costs to
economic agents, asset prices could not possibly reflect \emph{all}
information, so the hypothesis is usually interpreted in terms of
a normative pricing model. The EMH is an important stylisation for
constructing models of macroeconomic phenomena, for making investment
decisions and for assessing risk, and it has therefore been extensively
tested against data. However, the EMH has proved challenging to test
for at least two reasons. First, because we are faced with uncertainty
about the distribution of information among economic agents and the
costs they face in acting on this information. Secondly, since the
hypothesis is interpreted in terms of a normative pricing model, evidence
of a contradiction may be evidence against the pricing model only.

Models of asset prices are often required to show a so-called martingale
property, which roughly states that betting on prices should be a
fair game. Ever since Bachelier proposed a random walk model for stock
prices, researchers have been interested in characterising the intertemporal
dependence of the process generating returns. The independent increments
property of the random walk model is known to be inconsistent with
many stylised facts about stock returns: mean reversion and momentum
effects, predictability of returns at daily and weekly frequencies,
volatility clustering%
\footnote{In the finance literature, the term volatility is used quite specifically
to mean the statistical measure standard deviation, rather than other
types of variability. Volatility clustering refers to the observation
that the standard deviation of changes in asset prices is often high
for several consecutive periods, followed by several consecutive periods
of low volatility.%
} and higher volatility during U.S. market opening hours \citep{Fama1991}.
\citet{Fama1991} argues that predictable patterns in volatility of
stock price changes are not evidence against the EMH because such
predictability may be perfectly `rational'. General statistical dependence
in the process generating stock returns is not evidence against the
EMH, even though it implies that past returns can be informative about
the behaviour of future returns, because there is no necessary equilibrating
mechanism by which stock prices today should adjust to make our beliefs
about the pattern of future returns, which are conditional on past
returns, consistent with some unconditional beliefs. For example,
if today's stock prices reflect information about next year's stock
return volatility, this does not necessarily imply the existence of
arbitrage. Even though statistical dependence in the process generating
stock returns does not contradict the EMH, it has profound implications
for the proper specification of models to describe this process. 

The ideas of information redundancy and statistical dependence are
strongly related, because redundant information in a sample of stock
returns, for example, provides evidence for statistical dependence
in the latent model that generates those returns. This is the key
observation of our paper. Apart from being able to find substantial
information redundancy in stock returns at various frequencies and
in multiple stock markets, we are able to exploit this relationship
between redundancy and dependence, which the econometrics literature
has so far ignored to the best of our knowledge. Specifically we are
able to offer a statistical measure of and test for dependence. This
test for dependence can be applied to discover relationships in the
most general sense between different random structures -- even in
the presence of sampling variation -- and in particular can be applied
to discover intertemporal dependence in stock returns. In the category
of tests for dependence between real-valued random variables, including
tests of serial dependence, our test has some distinct advantages
over other existing methods. In general, our test can be used to discover
statistical dependence in any random structures that can be represented
in a computer, including text, pictures, audio and video.

Historically, measuring dependence in real-valued random variables
often amounted to measuring correlation between pairs of them. However,
correlation is a measure of linear association, which is only a specific
type of dependence. For example, if we consider a zero-mean symmetrically
distributed random variable $X$, then $X$ and $X^{2}$ have zero
correlation with each other even though they are completely dependent.%
\footnote{$X$ and $X^{2}$ have zero correlation because the covariance between
them is $Cov(X,X^{2})=\mathbb{E}X^{3}=0$. $X$ and $X^{2}$ are not
independent because $\sigma(X)\cap\sigma(X^{2})=\sigma(X)\neq\emptyset$.%
} Over and above this problem that occurs with a pair of variables,
correlation is unable to summarise the degree of dependence between
more than two variables in a single number. Correlation is therefore
not a sufficient or adequate measure of dependence. Many other measures
of dependence with desirable properties are available in econometrics,
including those based on entropy, mutual information, correlation
integrals, empirical distribution functions and empirical characteristic
functions. Our measure and test is most closely related to non-parametric
entropy-based dependence measures of \citet{Robinson1991}, \citet{Skaug1996},
\citet{Granger1994} and \citet{Hong2004}. These tests measure the
degree of dependence between pairs of random variables by estimating
a non-parametric (kernel) joint density function, computing its entropy,
and comparing this with the entropy of such a density under independence.
A large difference provides evidence against the independence of these
two random variables, and an asymptotic distribution theory allows
one to determine if the difference is larger than one might expect
as a result of sampling variation. Attractive features of entropy-based
tests are that they can identify non-linear dependence (unlike correlation)
and they are robust to monotonic transformations of the random variables.
The main disadvantages of the existing entropy-based tests cited above
are:
\begin{enumerate}
\item They have to be computed for pairs of random variables or pairs of
lags in a time series, so they cannot measure or be used to test for
joint dependence between sets of random variables. For example, they
cannot test whether lags 1 to 10 of a time series contain enough of
the intertemporal dependence that we may ignore other lags when specifying
a model.
\item They can be sensitive to the choice of bandwidth parameter and kernel
function made in density estimation%
. This problem is exacerbated by the fact that we need to estimate
a joint distribution, rather than just a single distribution.
\end{enumerate}
Our measure of and test for dependence overcomes both these problems,
while preserving the advantages of an entropy measure. It is a powerful
method that can test for intertemporal and cross-sectional dependence
in groups of time series. In fact, the full generality of the method
is remarkable.

The use of data compression in hypothesis testing is not novel.
It was introduced by \cite{ryabko2005application, ryabko2006universal, ryabko2006universal2}.
There, the authors propose, among others, a test for serial independence that requires the computation
of the so-called empirical entropy.
Our approach is different in that it does not rely on the empirical entropy.
In fact, our proposed test is very different and relies on computing \emph{several} compression ratios.
Loosely speaking, our test is as follows. 
Suppose you are given a sample from a stochastic process, $x_1,...,x_n$, and several transformations $T_0, T_1,...,T_m$ that can be applied to that sample. Suppose additionally that, under a null hypothesis $H_0$ to be tested, all the transformations $T_0(\{X_n\}), T_1(\{X_n\}), ..., T_m(\{X_n\})$ have the same distribution.
Then, under the null hypothesis, the compression ratios of the transformations, $\rho_0, \rho_1,...,\rho_m$ should be approximately the same, and we can reject the hypothesis if they are not.
In the case of our specific test all the transformations are based on random shuffles of the data, but the transformation $T_0$ is different from the others in that it shuffles `more' the data.
Then, we check whether the compression ratio corresponding to this shuffle $T_0$, $\rho_0$, is within a certain quantile of the remaining compression ratios, $\rho_1, ..., \rho_m$.

In the remainder of this introduction, we introduce the reader to
the relevant ideas in information theory and data compression that
motivate our test, and we describe the stock return data that will
be used illustratively throughout. In Section \ref{sub:Visualisation},
we offer a plotting procedure for visualising the degree of intertemporal
structure in a single time series. In Sections \ref{sub:Compression-ratios-and}
and \ref{sub:Introducing-the-lagged}, we apply our test of dependence
to identify serial dependence in several stock return series and known
stochastic processes. Section \ref{sec:Conclusion} concludes.

\subsection*{Acknowledgments}

The authors wish to thank Yoel Furman for fruitul discussions and insightful comments.

\subsection{Information and data compression \label{sub:Information-and-data}}

We can think of the \emph{information} contained in a set $\mathcal{A}$
as everything in $\mathcal{A}$ that is not redundant in some sense.
Compression can be achieved by finding redundancies and removing them.
Data compression is the process of finding small representations of
large quantities of data, where the size of the representation is
usually measured in bits. Compression can be \emph{lossless}, in which
case the small representation is exactly equivalent to the large representation,
or \emph{lossy}, in which case the small representation can be made
even smaller by also removing some of the most infrequently occurring
data atoms. In this paper we only need to consider lossless compression.

Consider the ordered sequence of real-valued random variables $\{X_{t}:t\geq0\}$
that we call the data generating process and suppose we have sampled
finitely many realisations of this process $\{x_{t_{i}}\}=\{x_{t_{1}},x_{t_{2}},x_{t_{3}},\ldots,x_{t_{n}}\}$
for $0\leq t_{1}<t_{2}<\ldots<t_{n}$.%
\footnote{We suppose that we have a filtered probability space on which this
random process is defined.%
} In order to represent real numbers in a computer, we need some discretisation
of the real line which defines the maximum and minimum observable
real numbers and the precision of the machine. In practice, we have
some discretion over this discretisation even if we are unaware of
it, and finer or wider partitions can be used when the application
requires it. The fineness of the partition is called the \emph{resolution}
of our machine, and this resolution will be relevant to our proposed
test for independence. We can therefore think of the sample space
of this stochastic process as being discrete. A \emph{code} for this
stochastic process is a bijection from the discretised real line to
the set of finite sequences of bits. Given a code, the original sequence
of symbols $\{x_{t_{i}}\}$ can be encoded into a sequence of bits,
which we call its \emph{bit representation}. The precise objective
of lossless compression is to find codes with the shortest possible
encoded sequences in the long run.

\begin{example} 
Consider a stochastic process $\{ X_n \}$ consisting of a collection of iid random variables with distribution $\bbP$ such that
\eqn{ \bbP(X = a) = \frac{1}{2}, \,\,\, \bbP(X = b) = \frac{1}{4}, \,\,\, \bbP(X = c) = \bbP(X = d) = \frac{1}{8}. }
In this case, one code achieving the best possible compression rate is given by \eqn{ a \to 0, \,\,\, b \to 10, \,\,\, c \to 110, \,\,\, d \to 111. } The average length of an optimally encoded sequence is, in the long run, $1.75 \, \mathrm{bits}/\mathrm{symbol}$. Note that this average length coincides with the entropy rate of the stochastic process, to be defined below.
\end{example}

Possibly the most important quantity in information theory is the
statistic known as \emph{entropy}. For a discrete random variable
$X_{1}$ with probability mass function $\{p_{1},p_{2},\ldots,p_{N}\}$,
the entropy is defined as%
\footnote{The logarithm is taken in basis $2$, as is is usual in Information
Theory. The same basis is used throughout the entire document even
though it is omitted.%
} 
\[
H(X_{1})=-\sum_{i=1}^{N}p_{i}\log(p_{i}),
\]
and for the unordered collection of random variables $X_{1},\ldots,X_{n}$
we can compute the entropy of the joint random variable $(X_{1},...,X_{n})$,
denoted $H(X_{1},...,X_{n})$. The entropy enjoys several remarkable
properties. One of these is that 
\[
H(X_{1},...,X_{n})\leq H(X_{1})+...+H(X_{n}),
\]
 with equality if and only if the random variables $X_{1},...,X_{n}$
are independent. For an overview of the remaining properties of entropy
we refer the reader to \citet{cover2006elements}. 

Given a stochastic process $\{X_{n}\}$, we can consider the entropy
of its first $n$ random variables and study how fast this quantity
grows. This leads to the definition of the \emph{entropy rate} of
the stochastic process $\{X_{n}\}$,
\[
h(\{X_{n}\})=\lim_{n}\frac{H(X_{1},...,X_{n})}{n}.
\]
whenever such a limit exists. It can be proven that for any stationary
process the entropy rate is well defined, see for instance \citep{cover2006elements}.
This entropy rate will form our measure of dependence and our test
statistic. Note that for a stationary process we have the inequality
\[
h(\{X_{n}\})\leq H(X_{1}),
\]
or, in other words,
\[
h(\{X_{n}\})=\alpha H(X_{1}),
\]
for some $\alpha\in[0,1].$ In other words, the entropy rate of a
stationary process can be decomposed in two components: the entropy
of its marginal distribution, $H(X_{1})$, and a discount factor reflecting
the intertemporal structure of the process, $\alpha$. In this work,
we propose to study the intertemporal structure of returns by analising
its entropy rate. For that purpose, we will need to control the contribution
of the entropy of the marginal distribution to the entropy rate -
we achieve this by a discretisation procedure described in detail
in Section \ref{sub:Visualisation}.

In the next two examples we contrast the behaviour of the entropy
rate under independence and dependence. %
The examples use restrictive assumptions to keep the exposition as
simple as possible.

\begin{example} Consider a stochastic process consisting of iid random variables, $\{ X_n \}$. Then, its entropy rate is \eqns{ h(\{ X_n \}) &= \lim_n \frac{H(X_1,...,X_n)}{n} \\ 									&= \lim_n \frac{\sum_{i = 1}^n H(X_i)}{n} \\ 									&= H(X_1). } 
That is, in the case of an iid sequence of random variables, the entropy rate od the stochastic process coincides with the entropy rate of its marginal distribution.
\end{example}
\begin{example} 
Consider a discrete one-step Markov chain with transition matrix $(P_{x_ix_j})$ for all $i,j$ and stationary distribution $\mu = (\mu_1,...,\mu_N)$. Let $\{ X_n \}$ be the stochastic process obtained when we start this Markov chain from its stationary distribution. Then 
\eqn{ \bbP(X_1 = x_1, ..., X_n = x_n) = \mu_{x_1} P_{x_1 x_2} ... P_{x_{n-1} x_n}, } and 
 \eqns{ \frac{1}{n} H(X_1,...,X_n) &= -\frac{1}{n} \sum_{(x_1,...,x_n)} \mu_{x_1} P_{x_1 x_2} ... P_{x_{n-1} x_n} \log(\mu_{x_1} P_{x_1 x_2} ... P_{x_{n-1} x_n}) \\ 					&= \frac{H(\mu) }{n} + \frac{(n-1)}{n} \sum_{(x_1,x_2)} \mu_{x_1} P_{x_1 x_2} \log(P_{x_1 x_2}) \\ 					&\overset{n}{\underset{\infty}{\to}} \sum_{x_1=1}^N \mu_{x_1} H(X_2 | X_1 = x_1), }
where $H(\mu)$ is the entropy of the stationary distribution. In other words, the entropy rate of a stationary one-step Markov chain is the average of the entropies of its $N$ conditional one-step random variables with weights given by the stationary distribution. 
In particular, the entropy rate of the stochastic process does not coincide with the entropy rate of its marginal distribution, it is strictly smaller.\footnote{In this case the entropy of the marginal distribution is $H(\mu)$.}. \end{example}

Intuitively, the entropy rate of a stochastic process is the rate
at which the number of `typical' sequences grows with the length of
the sequences. In other words, the number of typical sequences of
length $n$ for a given stochastic process $\{X_{n}\}$ is approximately
\[
2^{nh(\{X_{n}\})},
\]
for $n$ sufficiently large. It is essentially because of this interpretation
that, as we shall see, the entropy rate gives a sharp upper bound
to the maximum lossless compression ratio of messages generated by
a stochastic process. The \emph{Fundamental Theorem for a Noiseless
Channel} due to \citet{Shannon1948}, states that for any stationary
ergodic stochastic process $\{X_{n}\}$ and for any $\veps$, there
exists a code with average length smaller than $h(\{X_{n}\})+\veps$
bits/symbol. Furthermore, there does not exist any code with average
length smaller than $h(\{X_{n}\})$ bits/symbol.

Given a stochastic process $\{X_{n}\}$, if one has access to the
joint distribution of $(X_{1},...,X_{n})$, for $n$ sufficiently
large, then it is possible, at least in principle, to design codes
that are efficient for data generated by such a process. However,
in practice there are many situations in which one does not know beforehand
the distribution of the data generating process. Furthermore, even
in the limited number of cases in which one does know the distribution
of the data generating process, it might be unfeasible to compute
the joint distribution of $(X_{1},...,X_{n})$ for $n$ large, and/or
find the `typical' sequences for such a distribution. Therefore, it
is natural to ask whether there are universal procedures that will
effectively compress any data generated by a sufficiently regular
process. Our day-to-day experience with computers and celebrated tools
such as zip, tells us that the answer to this question should be positive.
And it is, indeed, as we shall see in the following. 

There is a popular class of compression algorithms that is universally
optimal and easy to implement. These are termed Lempel-Ziv, after
the authors of the two basic algorithms that underlie this class,
the LZ77 and LZ78. The LZ77 compression algorithm was introduced by
\citet{LZ77}. One year later the same authors published the LZ78
algorithm in \citet{LZ78}. Loosely speaking, the Lempel-Ziv algorithms
encode each sequence as a pointer%
\footnote{In computer science, a \emph{pointer} is an address to a location
in the computer's memory. It is an object, whose content is the address
of another object in the memory of a computer.%
} to the last time that sequence occurred in the data%
\footnote{More precisely, the algorithm stores new sequences in a dictionary
as they appear. This is why these algorithms are also described as
adaptive dictionary compression algorithms.%
}. Because typical sequences appear more often, the pointers to the
last time they occur are smaller than those for atypical sequences.
One remarkable feature of these algorithms, proved by Wyner and Ziv
\citeyearpar{LZ78proof,LZ77proof}, is that they are asymptotically
optimal. More precisely, given a stationary and ergodic stochastic
process $\{X_{n}\}$, and denoting by $l(X_{1},...,X_{n})$ the Lempel-Ziv
codeword length associated with $(X_{1},...,X_{n})$, 
\begin{equation}
\limsup_{n}\frac{l(X_{1},...,X_{n})}{n}\leq h(\{X_{n}\}),\,\,\, a.s..\label{eq:wz}
\end{equation}
 Since $1977$, improved implementations of Lempel-Ziv algorithms
have been developed. We use one of the most recent ones, the Lempel-Ziv-Markov
Algorithm (LZMA), which has been in development since $1998$ and
is featured in the widely available $.7z$ compression format. Combining
equation \ref{eq:wz} with the Fundamental Theorem for a Noiseless
Channel above, we can see that the average codeword length 
\[
\frac{l(X_{1},...,X_{n})}{n}
\]
is a consistent estimator for $h(\{X_{n}\})$ when the stochastic
process $\{X_{n}\}$ is stationary and ergodic. Given a computer file
containing an encoding of a random sample from the stochastic process
$\{X_{n}\}$, a proxy for $h(\{X_{n}\})$\emph{ }is its\emph{ compression
ratio }defined as 
\[
1-\frac{\mathrm{number\, of\, bits\, in\, the\, compressed\, file}}{\mathrm{number\, of\, bits\, in\, the\, uncompressed\, file}}.
\]
 More precisely, on the one hand, the compression ratio is an estimator
of the optimal compression ratio. On the other hand, the optimal compression
ratio, $\rho(\{X_{n}\})$, is related to the entropy rate via the
formula
\[
\rho(\{X_{n}\})=1-\frac{{h(\{X_{n}\})}}{\log(N)},
\]
where $N$ is the number of possible values that the process takes.
In the sequel, we will always quote compression ratios, instead of
entropy rates. The reader should recall the above formula to convert
between the two%
\footnote{Throughout this document we only compute the compression ratio of
stochastic processes taking exactly $256$ values. In this case, we
have $\log(256)=8$.%
}.

Unfortunately, to the best of our knowledge, there is currently no
general result for the rate of convergence of the Lempel-Ziv algorithms
to the entropy rate. To overcome this difficulty, we introduce extra
randomness in the data by randomly reshuffling blocks of data%
\footnote{By reshuffling we mean resampling without replacement.%
}. By comparing the compression rates obtained with and without the
extra randomness we are able to perform inference on the entropy rate
of the data-generating process. This procedure motivates the statistical
test for dependence that we describe in Section \ref{sub:A-summary-of}.

\subsection{Bias in the compression rate estimator}

The skeptical reader might be wondering how `optimal' these algorithms are, that is, how fast they approach optimality.
We end this section with a simple experiment that ought to `reassure' us that the algorithm does indeed work well.
Furthermore, as we sill see, this experiment teaches us that one more factor needs to be considered before we proceed with more serious matters.

\begin{example}[Convergence of the LZMA and overhead costs]
We start this example by generating a probability mass function with support in exactly $256$ points, that is a vector $p = (p_1,...,p_{256})$ such that $\sum p_i = 1$.
This vector of probabilities is generated randomly, but with the property that the first $128$ probabilities are on average smaller than the last ones\footnote{For the interested reader, we sampled $128$ uniform random variables in $[0,1/3]$ and $128$ uniform random variables in $[0,1]$. These $256$ samples are then used to form the probability vector - after being normalized so as to sum to $1$.}.
The entropy of the resulting distribution is readily computed as $7.484616$.

The next step in this experiment is to generate $N$ i.i.d. samples in ${0,...,255}$ according to the probability mass $p$.
Note that each sample can be perfectly represented using $1$ byte.
Shannon's fundamental theorem tells us, if $N$ is sufficiently large, then it should be possible to represent these $N$ samples using only $7.485$ bits, or $0.936$ bytes per sample.
In other words, for $N$ sufficiently large, there exist codes that achieve a compression ratio of approximately $0.064$.

To test how the LZMA algorithm performs on this compression task we write all the $N$ samples in a binary file and compress it.
Note that the size of the uncompressed file is exactly $N$ kilobytes.
The result of the compression is on the following table:

\begin{table}[!ht]
\centering
\begin{tabular}{rrrr}
  \hline
 & $N$ & compressed size (CS) & compression ratio (CR) \\ 
  \hline
1 & 100 & 222 & -1.220000 \\ 
  2 & 500 & 621 & -0.242000 \\ 
  3 & 1000 & 1109 & -0.109000 \\ 
  4 & 5000 & 5003 & -0.000600 \\ 
  5 & 10000 & 9874 & 0.012600 \\ 
  6 & 50000 & 48086 & 0.038280 \\ 
  7 & 100000 & 95598 & 0.044020 \\ 
  8 & 500000 & 475432 & 0.049136 \\ 
  9 & 1000000 & 950139 & 0.049861 \\ 
  10 & 5000000 & 4747766 & 0.050447 \\ 
  11 & 10000000 & 9495234 & 0.050477 \\ 
   \hline
\end{tabular}
\caption{Result of compressing data generated by a source with theoretical maximum compression ratio of $0.064$.}
\label{table:testEntropyTableShort}
\end{table}

Examining the table above generates mixed feelings. 
The compression ratio increases with $N$, as it should, but seems to staganate when it approaches $0.05$ which is not too close from the theoretical maximum.
This leaves us wondering about where our extra compression power went, and helps us realise that there is a factor that we are not taking into account.
The missing factor is \emph{overhead costs}, that is, costs inherent to how the algorithm is actually implemented.
These overhead costs can arise for a number of reasons, for example:
\begin{itemize}
\item The compressed file must carry a header with some information, which is a fixed cost.
\item The LZMA needs an initial dictionary to start compressing, which might be coded in the file at a fixed cost.
\end{itemize}
Helpfully, we can estimate this overhead cost and remove it, without guessing about its origins. 
Consider, for that purpose, a simple experiment analogous to the previous one. 
First, generate a sequence of $N$ independent and identically distributed random numbers uniformly distributed in $\{0,...,255\}$.
Then save the resulting sequence to a binary file and compress it using the LZMA algorithm as before. 
We know that this sequence is incompressible, that is its entropy $8$, the maximum possible.
Therefore, the resulting ``compressed'' file should have the overhead costs, but no actual compression.
The results of this second experiment can be found in Table \ref{table:OverheadCosts}.

\begin{table}[!ht]
\centering
\begin{tabular}{rrrrr}
  \hline
 & $N$ & ``compressed'' size & overhead & overhead (\%) \\ 
  \hline
1 & 100 & 223 & 123 & 1.230000 \\ 
  2 & 500 & 631 & 131 & 0.262000 \\ 
  3 & 1000 & 1135 & 135 & 0.135000 \\ 
  4 & 5000 & 5180 & 180 & 0.036000 \\ 
  5 & 10000 & 10242 & 242 & 0.024200 \\ 
  6 & 50000 & 50831 & 831 & 0.016620 \\ 
  7 & 100000 & 101488 & 1488 & 0.014880 \\ 
  8 & 500000 & 506982 & 6982 & 0.013964 \\ 
  9 & 1000000 & 1013758 & 13758 & 0.013758 \\ 
  10 & 5000000 & 5068105 & 68105 & 0.013621 \\ 
  11 & 10000000 & 10135980 & 135980 & 0.013598 \\ 
   \hline
\end{tabular}
\caption{Result of compressing data generated by a source with theoretical maximum compression ratio of $0$.}
\label{table:OverheadCosts}
\end{table}

As we can see from this table, the overhead costs decrease with the length of our sample $N$, but is nevertheless non-negligible.
These overhead costs bias our estimator of the optimal compression ration.
If one takes into account these bias, we can recast the results in Table \ref{table:testEntropyTable}.

\begin{table}[!ht]
\centering
\begin{tabular}{rrrrrr}
  \hline
 & $N$ & CS & CS minus overhead & `unbiased' CR \\ 
  \hline
1 & 100 & 222 & 99 & 0.010000 \\ 
  2 & 500 & 621 & 490 & 0.020000 \\ 
  3 & 1000 & 1109  & 974 & 0.026000 \\ 
  4 & 5000 & 5003  & 4823 & 0.035400 \\ 
  5 & 10000 & 9874 & 9632 & 0.036800 \\ 
  6 & 50000 & 48086  & 47255 & 0.054900 \\ 
  7 & 100000 & 95598  & 94110 & 0.058900 \\ 
  8 & 500000 & 475432  & 468450 & 0.063100 \\ 
  9 & 1000000 & 950139  & 936381 & 0.063619 \\ 
  10 & 5000000 & 4747766  & 4679661 & 0.064068 \\ 
  11 & 10000000 & 9495234  & 9359254 & 0.064075 \\ 
   \hline
\end{tabular}
\caption{Result of compressing data generated by a source with theoretical maximum compression ratio of $0.064$ taking into account overhead costs.}
\label{table:testEntropyTable}
\end{table}

Once the overhead costs are removed, we can see the compression ratio converging to the theoretical maximum.
In fact, for sequences of size $5e5$ the compression ratio is already close to being optimal.
Most of the datasets we will use in the sequel are smaller than this.
However, the dataset of S\&P500 tick data is precisely of this size.
Indeed, high frequency data seems to be a good target for these techniques.
\end{example}

The previous example shows that our estimator of the optimal compression ratio has a bias due to overhead costs.
It also shows that it is possible to remove this bias by estimating the overhead costs using an uncompressible source of data.
In what follows, we will account for this bias and report only bias-corrected estimates of the compression ratios.

\subsection{The stock return data}

The two stock indices used in this paper are the S\&P500 and the headline
index of the Portuguese Stock Exchange, the PSI20. Various sample
periods are used and sampling frequencies of day, minute and tick
are considered. We summarise the returns series over the various samples
in Table \ref{tab:Summary-statistics}. All data were obtained from
Bloomberg%
\footnote{Tickers: ``SPX Index'' and ``PSI20 Index''.%
}. 
\begin{table}[h]
\caption{Summary statistics for the return series data used in this paper.
\label{tab:Summary-statistics}}

\begin{centering}
\begin{tabular}{rlll}
  \hline
 & start & end & medianFreq \\ 
  \hline
SPdaily & 1928-01-03 & 2013-02-28 & 1 days \\ 
  SPdaily1 & 1928-01-03 & 1949-02-10 & 1 days \\ 
  SPdaily2 & 1992-12-03 & 2013-02-28 & 1 days \\ 
  SPmin & 2012-08-24 & 2013-03-01 & 1 mins \\ 
  SPtick & 2012-08-24 & 2013-03-01 & 5 secs \\ 
  PSI20daily & 1993-01-04 & 2013-02-28 & 1 days \\ 
  PSI20min & 2012-08-24 & 2013-03-01 & 1 mins \\ 
  PSI20tick & 2012-08-24 & 2013-03-01 & 15 secs \\ 
   \hline
\end{tabular}
 
\par\end{centering}

\begin{centering}
\medskip{}

\par\end{centering}

\centering{}\resizebox{\textwidth}{!}{
\begin{tabular}{rrrrrrrrrr}
  \hline
 & 0\% & 25\% & 50\% & 75\% & 100\% & mean & std & skew & kurtosis \\ 
  \hline
SPdaily & -0.2290 & -0.004728 & 0.0006116 & 0.005506 & 0.1537 & 0.00021137 & 0.0120193 & -0.43 & 18.5 \\ 
  SPdaily1 & -0.1386 & -0.007055 & 0.0008227 & 0.007590 & 0.1537 & -0.00003762 & 0.0172679 & 0.02 & 8.0 \\ 
  SPdaily2 & -0.0947 & -0.004894 & 0.0005637 & 0.005848 & 0.1096 & 0.00024832 & 0.0120808 & -0.24 & 8.4 \\ 
  SPmin & -0.0091 & -0.000123 & 0.0000066 & 0.000124 & 0.0122 & 0.00000150 & 0.0003104 & 1.15 & 123.3 \\ 
  SPtick & -0.0036 & -0.000020 & -0.0000065 & 0.000020 & 0.0065 & 0.00000014 & 0.0000622 & 1.65 & 487.2 \\ 
  PSI20daily & -0.1038 & -0.004769 & 0.0003603 & 0.005537 & 0.1020 & 0.00013639 & 0.0114057 & -0.35 & 8.5 \\ 
  PSI20min & -0.0191 & -0.000183 & 0.0000038 & 0.000188 & 0.0122 & 0.00000290 & 0.0004063 & -0.88 & 158.7 \\ 
  PSI20tick & -0.0189 & -0.000096 & 0.0000034 & 0.000099 & 0.0116 & 0.00000102 & 0.0002448 & -1.92 & 363.5 \\ 
   \hline
\end{tabular}
}
\end{table}

\section{Method and results}

In this section, we describe our methodology and present our results.
In Section \ref{sub:Visualisation}, we describe how an effective
discretisation of the data is useful to detect its intertemporal statistical
structure. In Section \ref{sub:Compression-ratios-and}, we describe
how combining a bootstrapping method with compression algorithms can
give insight on the dependence of random variables. In the same section,
we present our findings in the returns data. In Section \ref{sub:Introducing-the-lagged},
we introduce the serial dependence function which measures the increase
in dependence obtained from considering bigger collections of consecutive
time-series points. In Section \ref{sub:A-summary-of}, we summarise
the methodology by formalising it a statistical test. Finally, in
Section \ref{sub:Further-examples}, we test our methodology in synthetic
data. Additionally, in the same section, we use this technique to
measure the goodness of fit of a GARCH(1,1) model to S\&P500 daily
returns.

\subsection{Visualisation of temporal dependence in quantiles at high frequencies
\label{sub:Visualisation}}

Consider again the sample $\{x_{t_{i}}\}=\{x_{t_{1}},x_{t_{2}},x_{t_{3}},\ldots,x_{t_{n}}\}$
for $0\leq t_{1}<t_{2}<\ldots<t_{n}$ defined in Section \ref{sub:Information-and-data}.
We want to study the `structure' of this sample to make inferences
about the data generating process. The structure that we examine is
the relationship between the longitudinal or time series ordering
of the sample to its ordering in terms of the $\leq$ relation on
$\mathbb{R}$. We could define the (longitudinal) empirical cumulative
distribution function%
\footnote{If the stochastic process generating the data is stationary and ergodic
then, by the ergodic theorem, the longitudinal empirical cumulative
distribution function converges to the marginal cumulative distribution
function.%
} 
\begin{equation}
\hat{F}(x)=\frac{1}{n}\sum_{i=1}^{n}\mathbb{I}\{x_{t_{i}}\leq x\},\label{eq:cdf}
\end{equation}
and by applying the transformation $x\longmapsto n\hat{F}(x)$ to
our sample, we would obtain the time-ordered ranks of the values in
our sample.
These ranks are detailed in that they span the integers $\{1,2,\ldots,n\}$,
and for the large samples that we consider, the detail prevents us
from seeing patterns in our sample that are present across time with
the naked eye. Now suppose that we group the observations into $2^{8}=256$
equally-sized `bins' based on their rank, using the transformation
$x\longmapsto\lfloor2^{8}\hat{F}(x)\rfloor$, where $\lfloor\cdot\rfloor$
is the usual floor function. Doing so would reduce the detail in the
rank-ordering of our sample, which can be seen as a decrease in resolution
or a loss of information. We would obtain from our sample a sequence
of length $n$, taking values on the integers $\{1,2,3,\ldots,2^{8}\}$,
and this sequence would describe the time evolution of high and low
values in our sample, where `high' and `low' are understood in terms
of the set $\{1,2,3,\ldots,2^{8}\}$ rather than the set $\{1,2,\ldots,n\}$.
To enable a visualisation, we have effectively discretised the sample
space $\mathbb{R}$ into $2^{8}$ regions based on the quantiles $\{2^{-8}i:i=0,1,2,\ldots,2^{8}\}$
of the distribution function $\hat{F}$. The exact partition of $\mathbb{R}$
obtained via this method is the one obtained by applying the inverse
cumulative distribution function
\[
\hat{F}^{-1}(q):=\inf\{x|\hat{F}(x)\leq q\}
\]
to $[0,1]$. Therefore, our discretisation will have more resolution
in areas where the density corresponding to $\hat{F}$ is more peaked.
We could use the terminology `states' to refer to the indexing set
$\{1,2,3,\ldots,2^{8}\}$ and we could plot these states against time
to visualise local serial dependencies in the value of our process.
If the underlying data generating process were a sequence of independent
(but not necessarily identically distributed) random variables, we
would expect a uniformly random scatter of points across states and
time. We say that this procedure performs a \emph{rank plot} with
8 bits of resolution.

\begin{figure}
\begin{centering}
\vspace{-1.5cm}\subfloat[Day]{\begin{centering}
\includegraphics[scale=0.35]{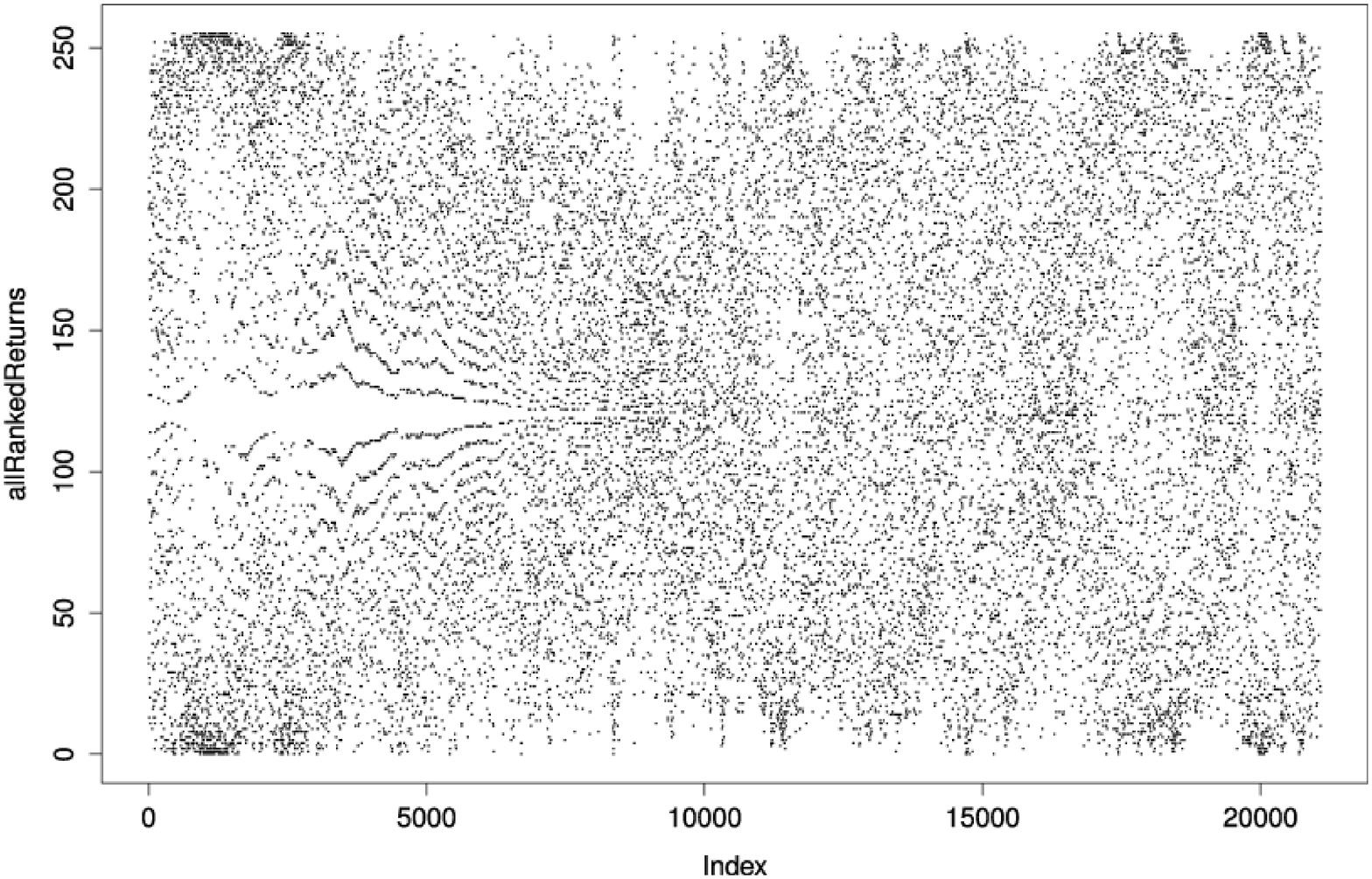}
\par\end{centering}

}
\par\end{centering}

\begin{centering}
\subfloat[Minute]{\begin{centering}
\includegraphics[scale=0.35]{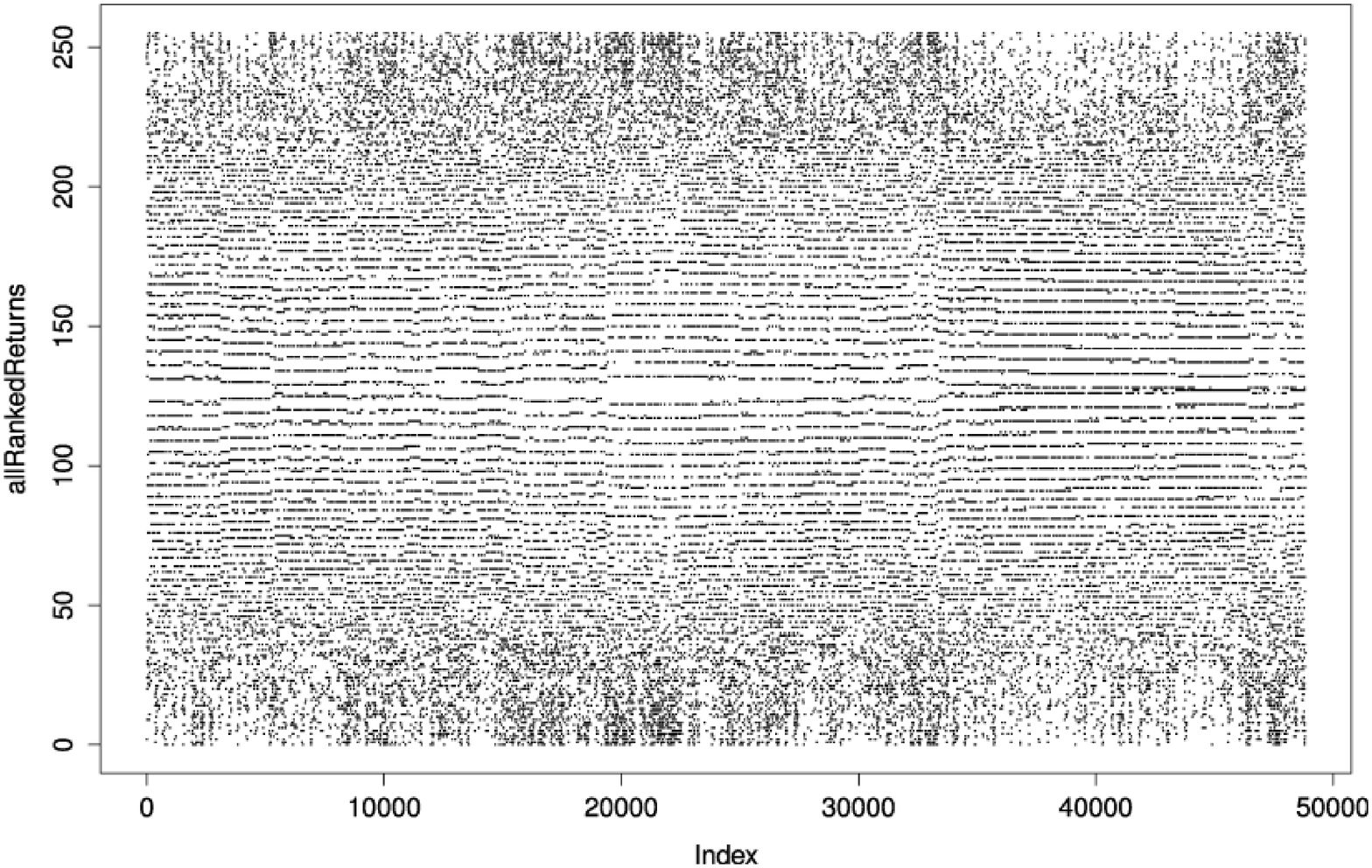}
\par\end{centering}

}
\par\end{centering}

\begin{centering}
\subfloat[Tick]{\begin{centering}
\includegraphics[scale=0.35]{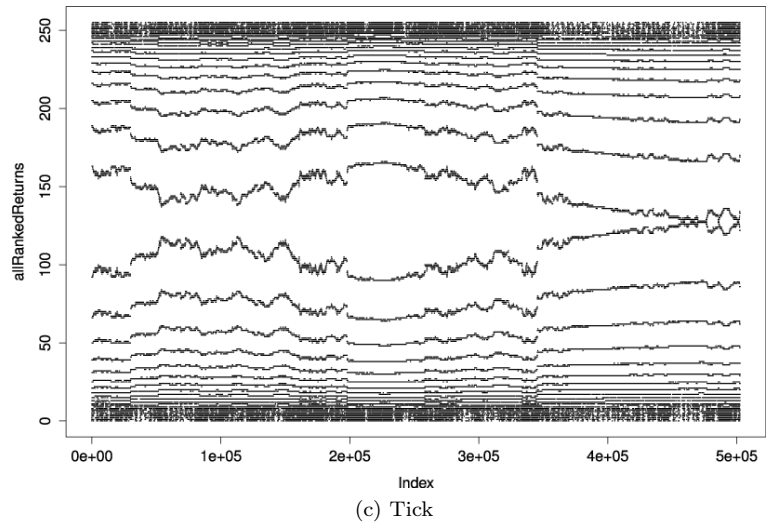}
\par\end{centering}

}
\par\end{centering}

\centering{}\caption{The time evolution of $256$ quantiles of returns at daily, hourly
and tick frequencies. \label{fig:rank-time-plot}}
\end{figure}

We apply the above method to our samples of S\&P500 returns at day,
minute and tick frequencies and plot the results in Figure \ref{fig:rank-time-plot}.
Note that each panel in Figure \ref{fig:rank-time-plot} shows only
one time series, despite the discernible horizontal lines. In this
rank plot with 8 bits of resolution, we can interpret the returns
series as a discrete Markov chain of unknown lag order, transitioning
at every point on the $x$-axis to one of $2^{8}$ possible states.
From inspecting the figure, we can discern some time structure in
the returns, in the form of patterns in the black and white regions
of the plot, with more structure at higher frequencies. The daily
returns show more time structure in the first half of the sample period
than in the second half. Near the beginning of the sample period,
the observations are clustered in the top and bottom states, which
are the extremes of the (longitudinal) distribution of returns, and
clustered away from the median returns. The presence of horizontal
patterns in the first half of the daily returns sample period indicates
some clustering of similar states or quantiles in small time intervals.

Smart algorithms for pattern detection, like data compression algorithms,
can discover these patterns and others hidden from the human eye.
By representing the original data equivalently as combinations of
commonly-occuring patterns, these algorithms can distill a large data
set into an equivalent smaller data set, and the amount by which the
size of the data set is reduced is a measure of the redundancy in
the original data set.

\subsection{Compression ratios and testing for serial dependence \label{sub:Compression-ratios-and}}

Returns in a computer are most often represented as 32- or 64-bit
numbers.
We standardise the amount of memory allocated to each return by representing
each return using only 8 bits, which effectively applies the transformation
$x\longmapsto\lfloor2^{8}\hat{F}(x)\rfloor$ of the previous section
to each return. Higher resolutions of 32 and 64 bits are also possible,
and would result in higher compression ratios, without affecting the
validity of the statistical test we would like to propose. We know
that returns are not likely to be any number that a machine can represent
as a 64-bit number. For example, we know the we will never observe
a return of $2^{60}$, even though by representing numbers as doubles,
we are effectively considering that possibility as well. Furthermore,
the returns have a specific distribution which over time is centered
and concentrated around zero. Just by exploiting this fact we can
get a good compression ratio. However, this gain does not have to
do with the intertemporal statistical structure of the signal, which
is precisely what we propose to study. Therefore, we transform the
returns in the above way so that all combinations of 8 bits appear
equally often. These issues of the resolution with which to represent
numbers are implicitly present also in other econometric tests of
dependence. 

After choosing the resolution, the various redundancies that a lossless
compression algorithm can detect are
\begin{enumerate}
\item Time series patterns -- this is the temporal dependency that we would
like to isolate from the entropy rate estimator of the returns series.
\item Within-number patterns -- these are inefficiencies in the machine
representation of the numbers themselves and would arise, for example,
if we used a 64 bit number to represent a return with mean close to
zero and standard deviation $\ll1$.
\end{enumerate}
By applying our above transformation and 8-bit representation, we
can eliminate much of the non-temporal relationships in our data.
In other words, we control the contribution of the marginal distribution
of returns to the entropy rate of the return process. Then, to study
the intertemporal component of the entropy rate we perform random
shuffles of blocks of data. By varying the block size used in the
shuffling procedure, we can assess the contribution of the dependence
within each block to the overall entropy rate of the process. Note
that in the extremes of this spectrum of shuffles are the shuffles
of blocks of size $1$ - which shuffles all the data and destroys
all the intertemporal structure - and the shuffles of blocks of size
$n$ - which does not shuffle at all and therefore does not destroy
any intertemporal structure.

\begin{figure}
\vspace{-1.8cm}
\hspace*{-1.7cm}\subfloat[S\&P500 Day]{\includegraphics[scale=0.32]{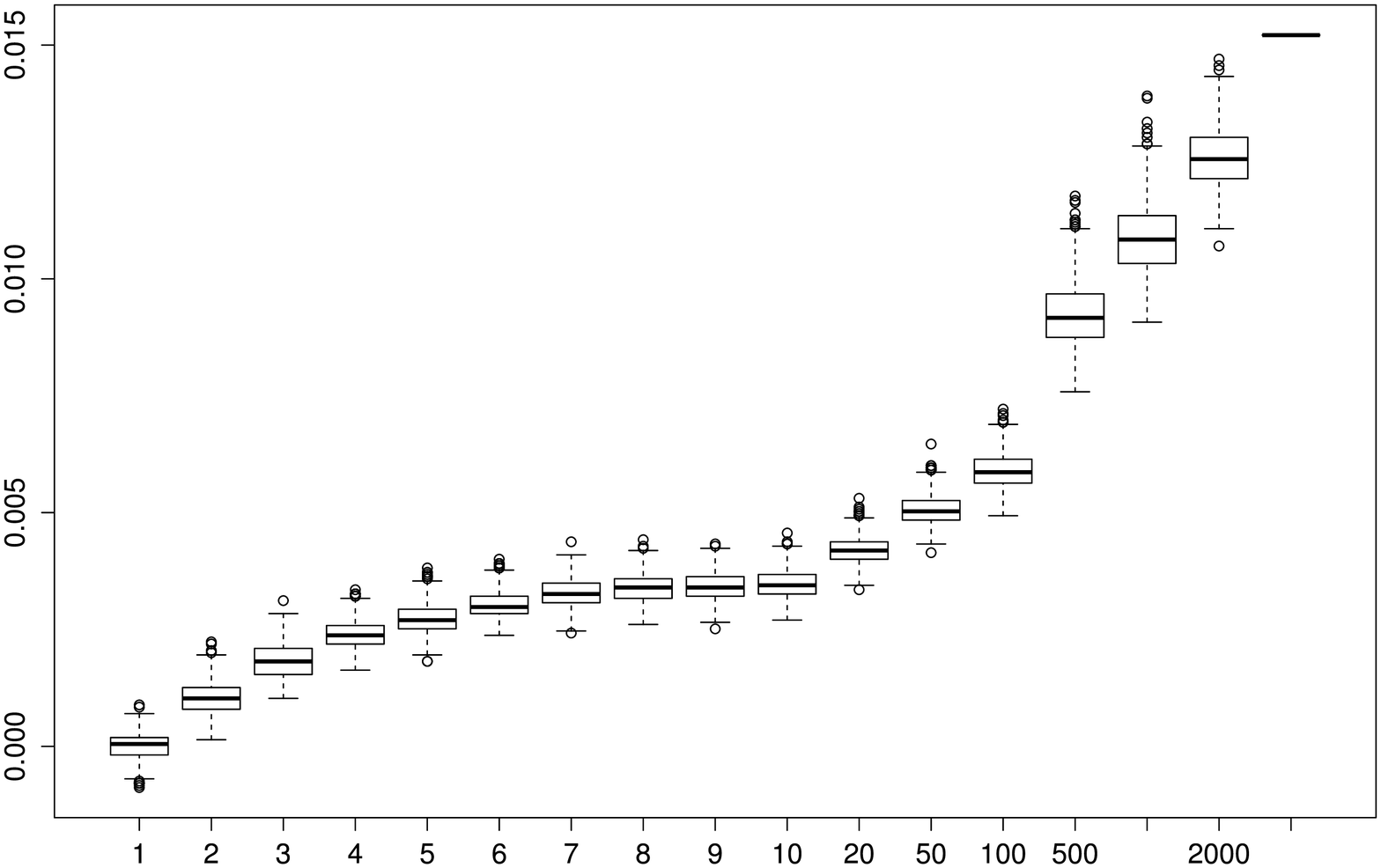}

}\hspace{1cm}\subfloat[PSI Day]{\includegraphics[scale=0.32]{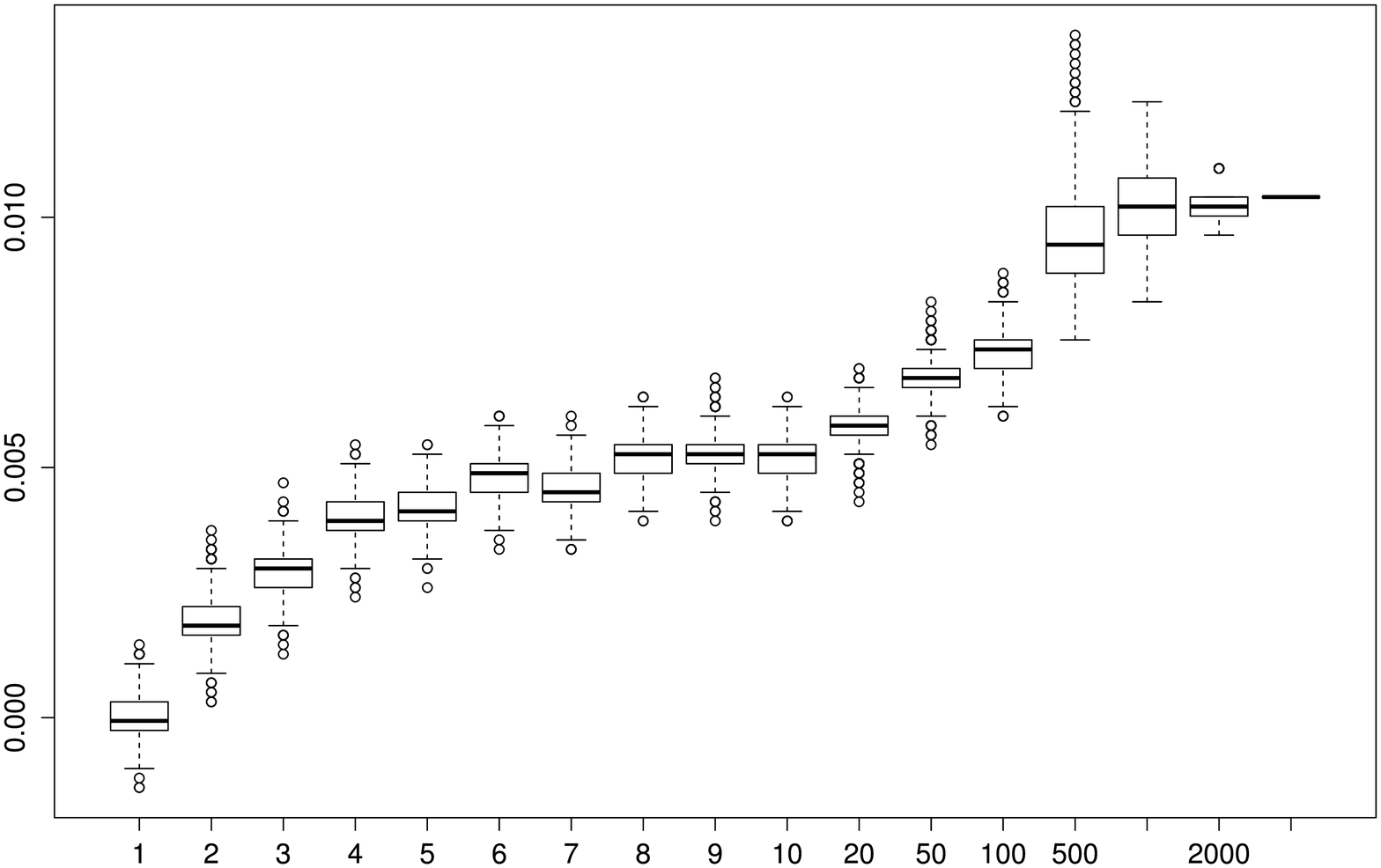}

}
\vspace{1cm}
\hspace*{-1.7cm}\subfloat[S\&P500 Minute]{\includegraphics[scale=0.32]{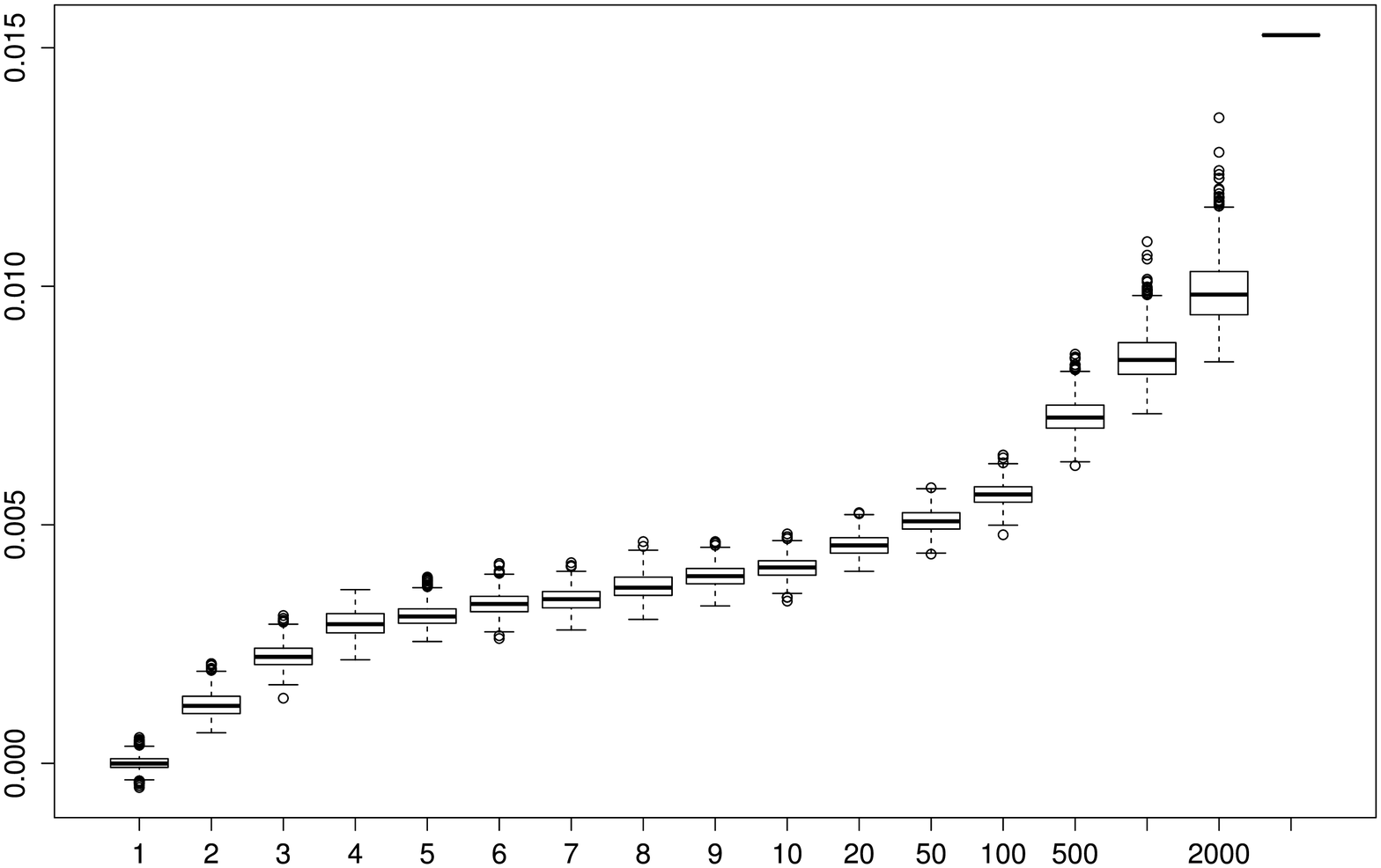}

}\hspace{1cm}\subfloat[PSI Minute]{\includegraphics[scale=0.32]{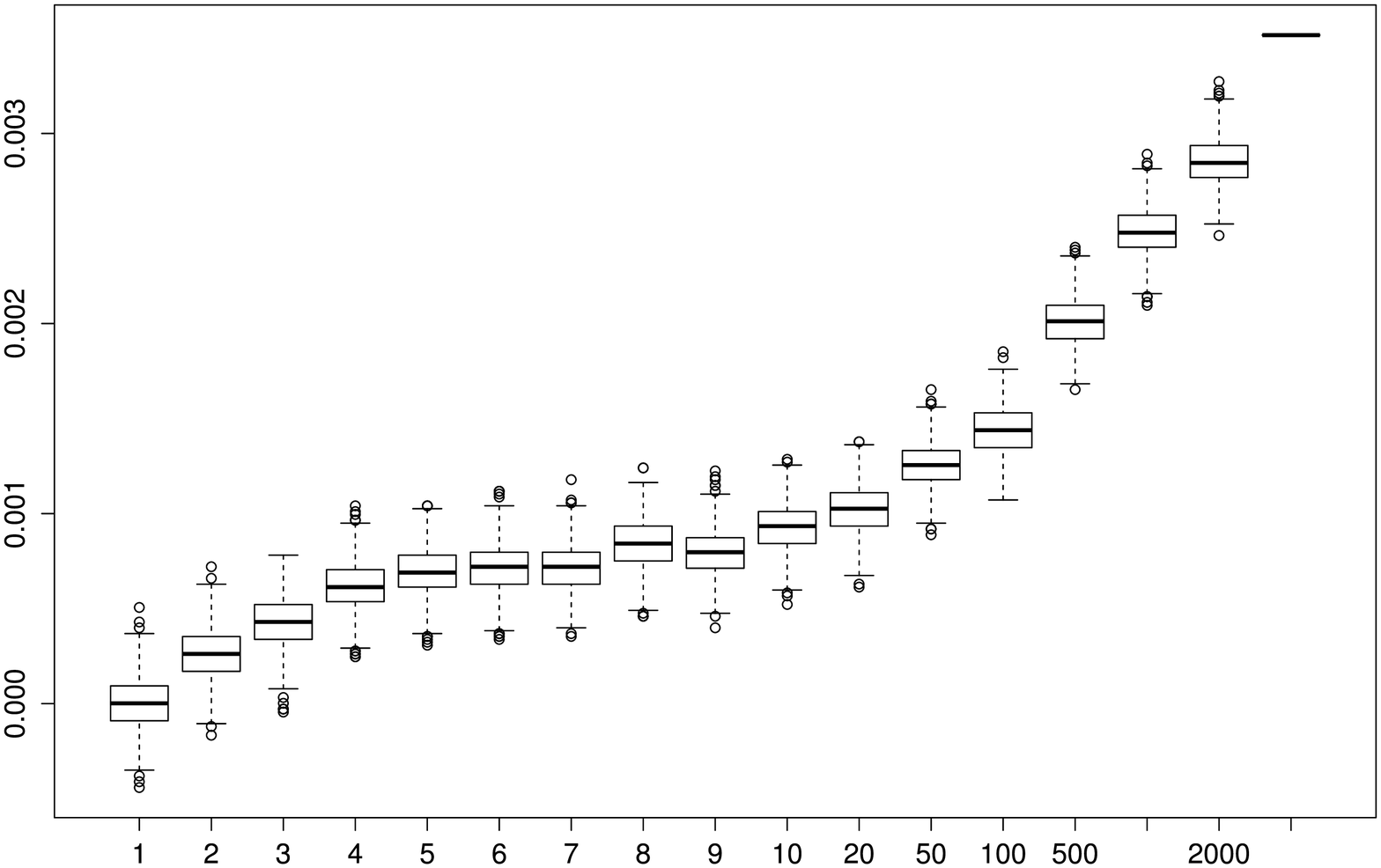}

}
\vspace{1cm}
\hspace*{-1.7cm}\subfloat[S\&P500 Tick]{\includegraphics[scale=0.32]{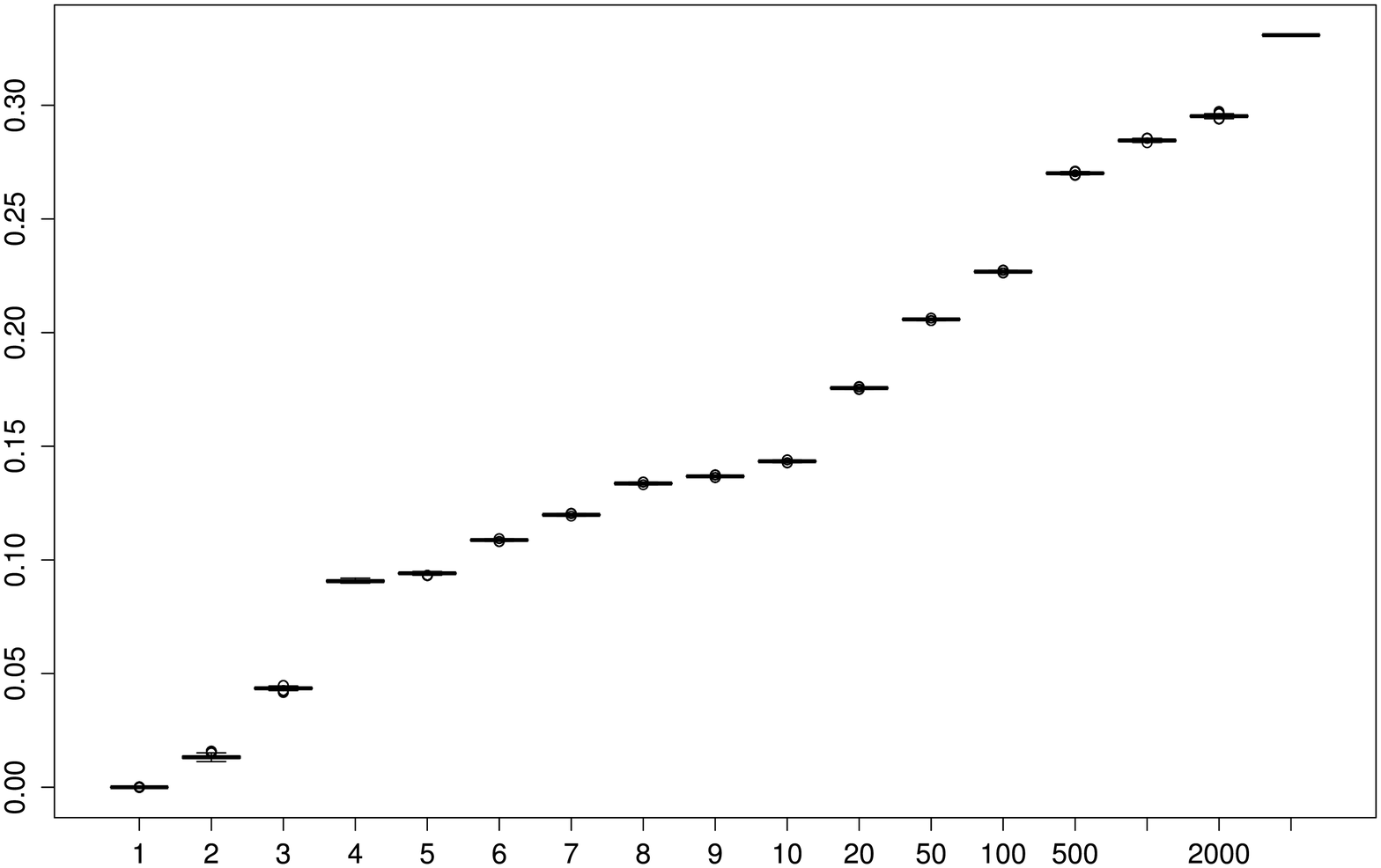}

}\hspace{1cm}\subfloat[PSI Tick]{\includegraphics[scale=0.32]{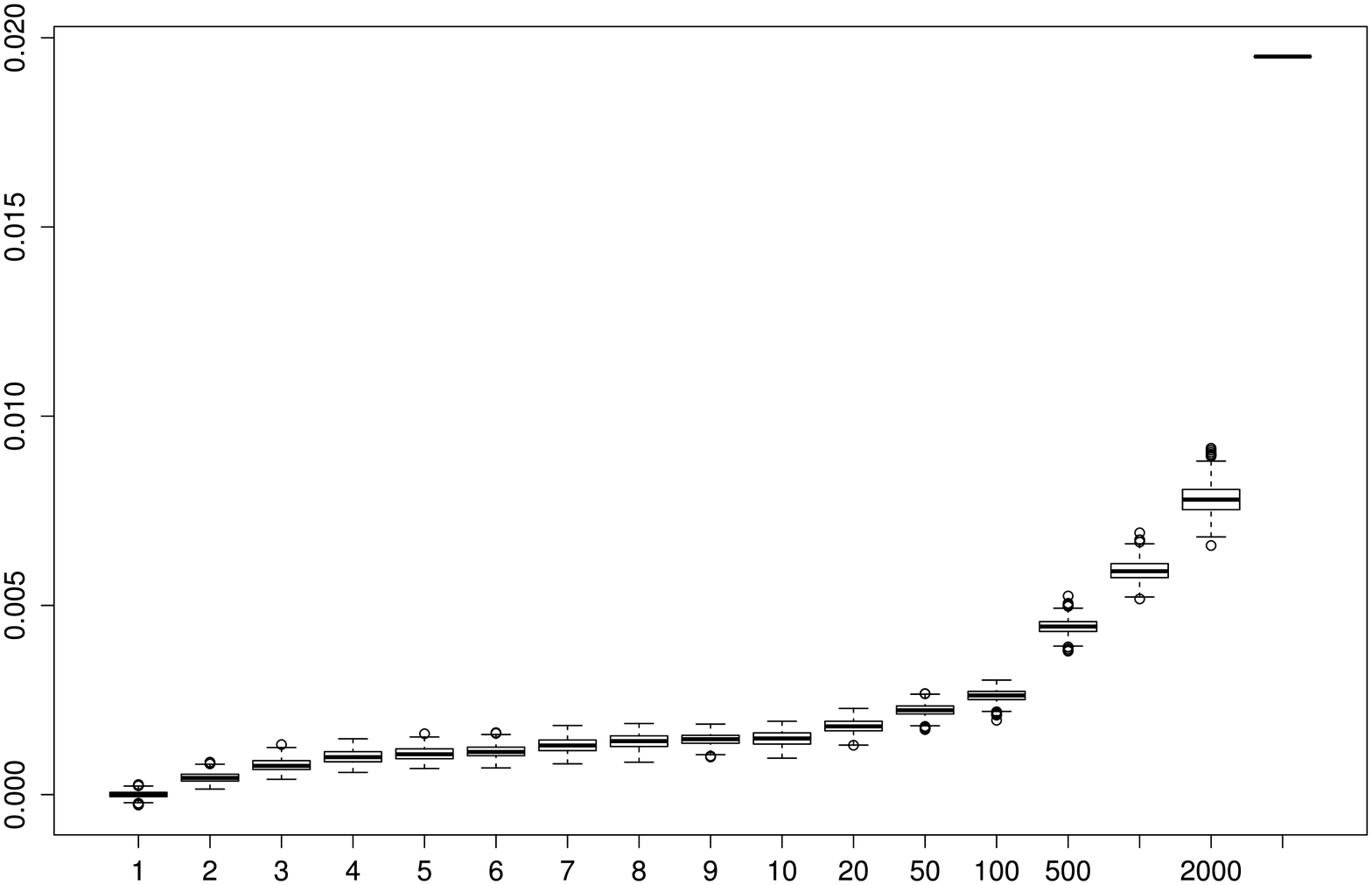}

}

\caption{Entropy rate estimators for original return series, and box-and-whisker
plots summarising the entropy rate estimators for shuffled return
series at various block sizes. Consecutive returns are resampled in
blocks of the specified size (without replacement) and for each shuffled
time series of returns a compression rate estimate is calculated.
\label{fig:Compression-ratios}}
\end{figure}

In Figure \ref{fig:Compression-ratios}, we show the compression ratios
achievable for the S\&P500 and PSI returns series at frequencies of
day, minute and tick for. The $y$-axis plots the unbiased estimate
of the compression ratio of the underlying stochastic process computed
for shuffles of blocks of varying length (ranging from $1$ to $n$,
where $n$ is the length of the time-series). In each panel, the estimate
of the compression ratio of the returns generating process is the
vertical height at the rightmost point on the $x$-axis. A higher
compression ratio indicates more statistical redundancy in the returns
series, which occurs at higher frequencies and more so on the PSI
than the S\&P500. 

By randomly shuffling the individual returns before compressing, we
are able to destroy all statistical structure/patterns in the time
dimension, so that all remaining compression would come from inefficiencies
in the machine representation of the numbers, less the overhead cost.
Note that our 8 bit representation is very efficient so that the compressibility
of the file is close to zero. The leftmost point on the $x$-axis
in each panel of Figure \ref{fig:Compression-ratios} gives our compression
rate estimate of an independent and identically uniform random variable,
which has a mean of exactly zero and some (small) uncertainty from
sampling variation.
The horizontal axis in each panel shows the block size used to perform
the random shuffling. For each block size we perform $1000$ shuffles
so as to be able to compute confidence intervals. Returns series that
have been shuffled in block sizes of 1 seem to be effectively iid
and achieve a much lower compression than the original (unshuffled)
returns series, which can be interpreted as saying that there is more
redundancy/less statistical information in the original returns. Equivalently,
the result can be interpreted as evidence for the informativeness
of the time series dimension in returns data. If returns were independent
(but not necessarily identically distributed) across time, their information
content would not be affected by shuffling. By comparing the estimated
compression ratio for the original time-series and the distribution
of estimated compression rates obtained from many randomly shuffled
returns, it is possible to conduct a statistical test of the independence
assumption.%
\footnote{Each shuffle of the returns is random and is independent of the previous
shuffle. We effectively sample with replacement from the $n!$ permutations
of a vector of a length $n$.%
} In every panel, we see that the $p$-value of such a test would be
zero, in the sense that the compression ratio from the original returns
series lies above the maximum compression ratio of the returns series
shuffled in blocks of size 1.

By increasing the block size, we destroy less time series structure
in the original returns series, so that more compression is, on average,
achievable. Equivalently, by increasing the block size, we provide
more structured pseudo-return data to the compression algorithm, which
tries to detect and exploit any extra information. Even at block sizes
of 10 consecutive returns, we note that the compression ratio of the
original return series lies above the maximum compression ratio of
the shuffled return series, which means that a significant proportion
of the dependence that is detectable in the original returns series
is due to dependence at lags greater than 10.

As the block size grows, the compression rate estimate of the shuffled
returns approaches the compression ratio for the original returns
series. For any given tolerance, there is a cutoff point defined by
a statistical test where the modeller is indifferent between models
of lag larger than $k$ - this lag would be the one that captures,
within the given tolerance, the intertemporal statistical structure
of the whole series. For the block sizes shown in Figure \ref{fig:Compression-ratios},
we can only identify this cutoff for the daily PSI returns. For a
typical tolerance level in $[0,1/2)$, we expect the cutoff to be
below lag 499, and closer to 499 than 99. In other words, for the
PSI daily returns and for a tolerance level in $[0,1/2)$ a modeller
will chose models with lags smaller than 499.%
{} It is then natural to consider the rate at which the mean compression
ratio grows with the lag length or block size.

\subsection{The lagged dependence measure and the compression ratio gap \label{sub:Introducing-the-lagged}}

\begin{figure}
\vspace{-1.8cm}
\hspace*{-1.7cm}\subfloat[S\&P500 Day]{\includegraphics[scale=0.32]{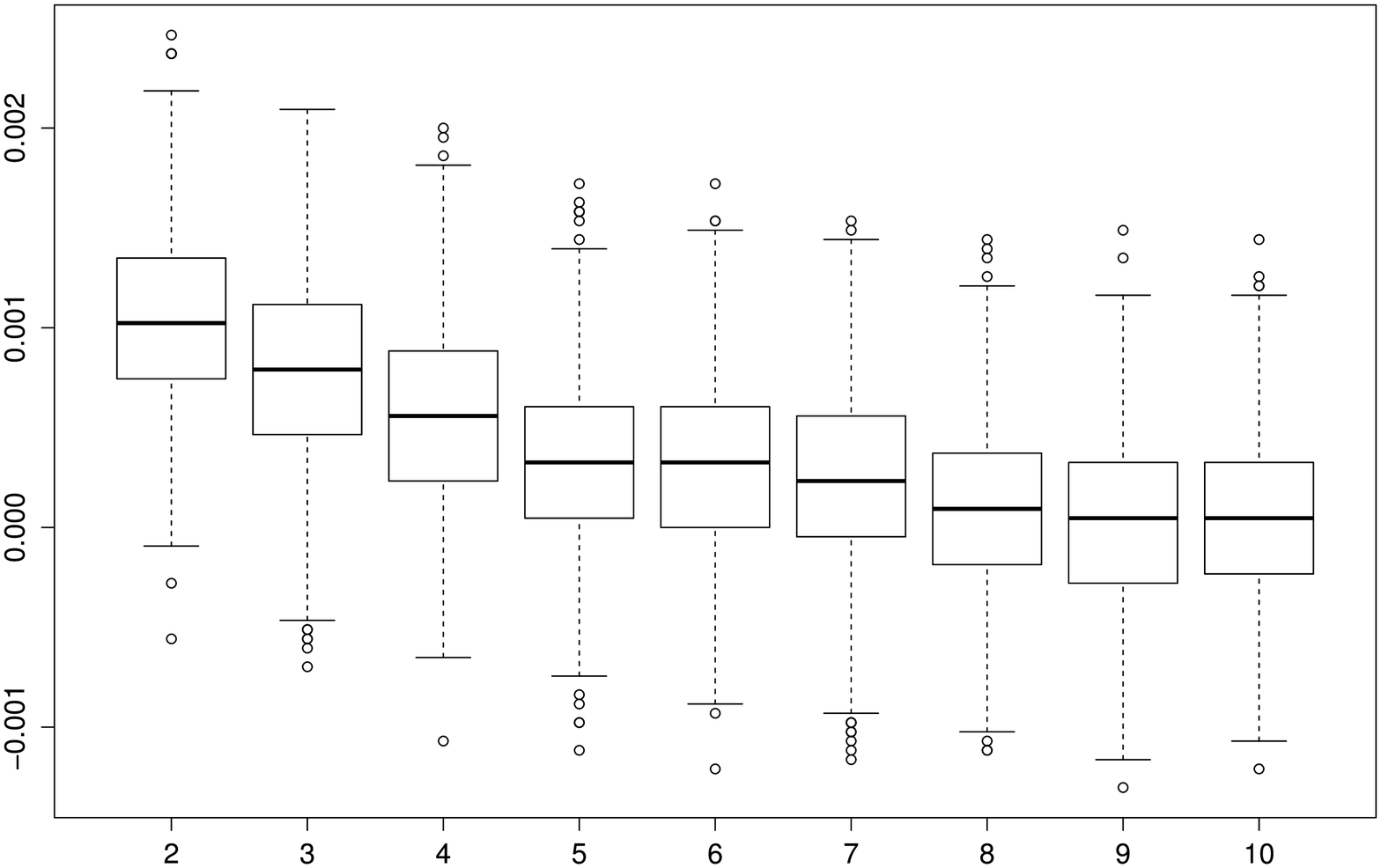}

}\hspace{1cm}\subfloat[PSI Day]{\includegraphics[scale=0.32]{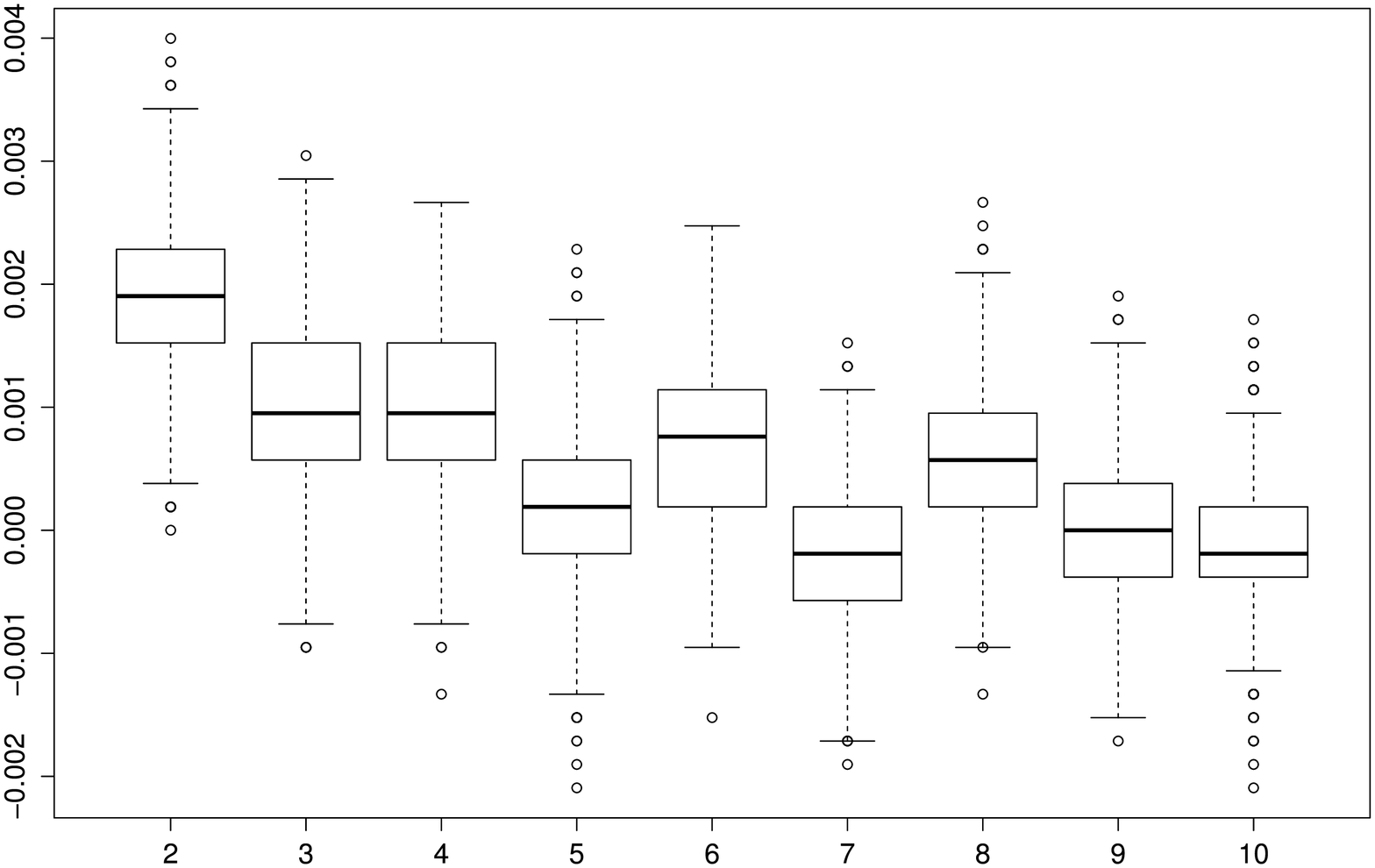}

}
\vspace{1cm}
\hspace*{-1.7cm}\subfloat[S\&P500 Minute]{\includegraphics[scale=0.32]{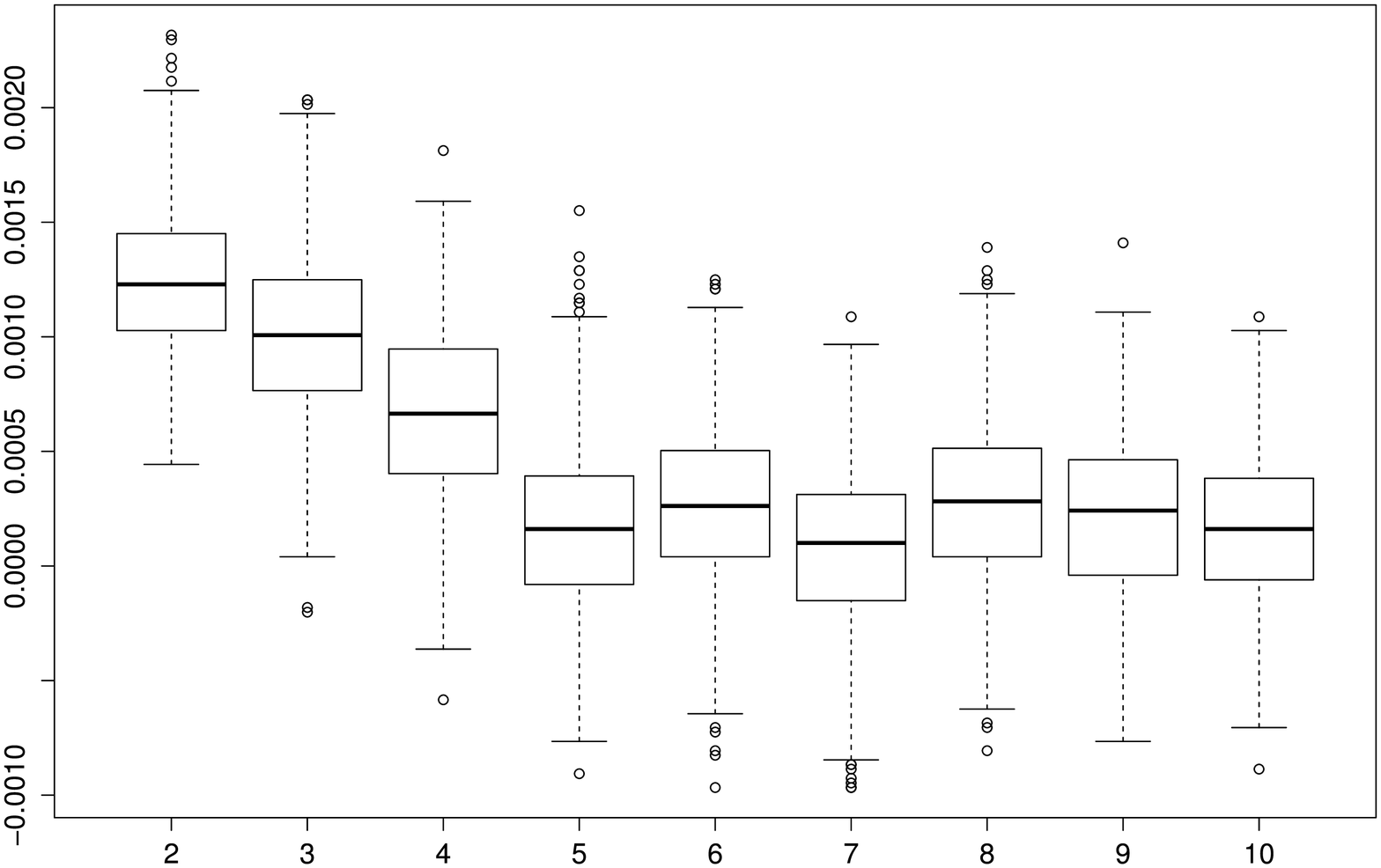}

}\hspace{1cm}\subfloat[PSI Minute]{\includegraphics[scale=0.32]{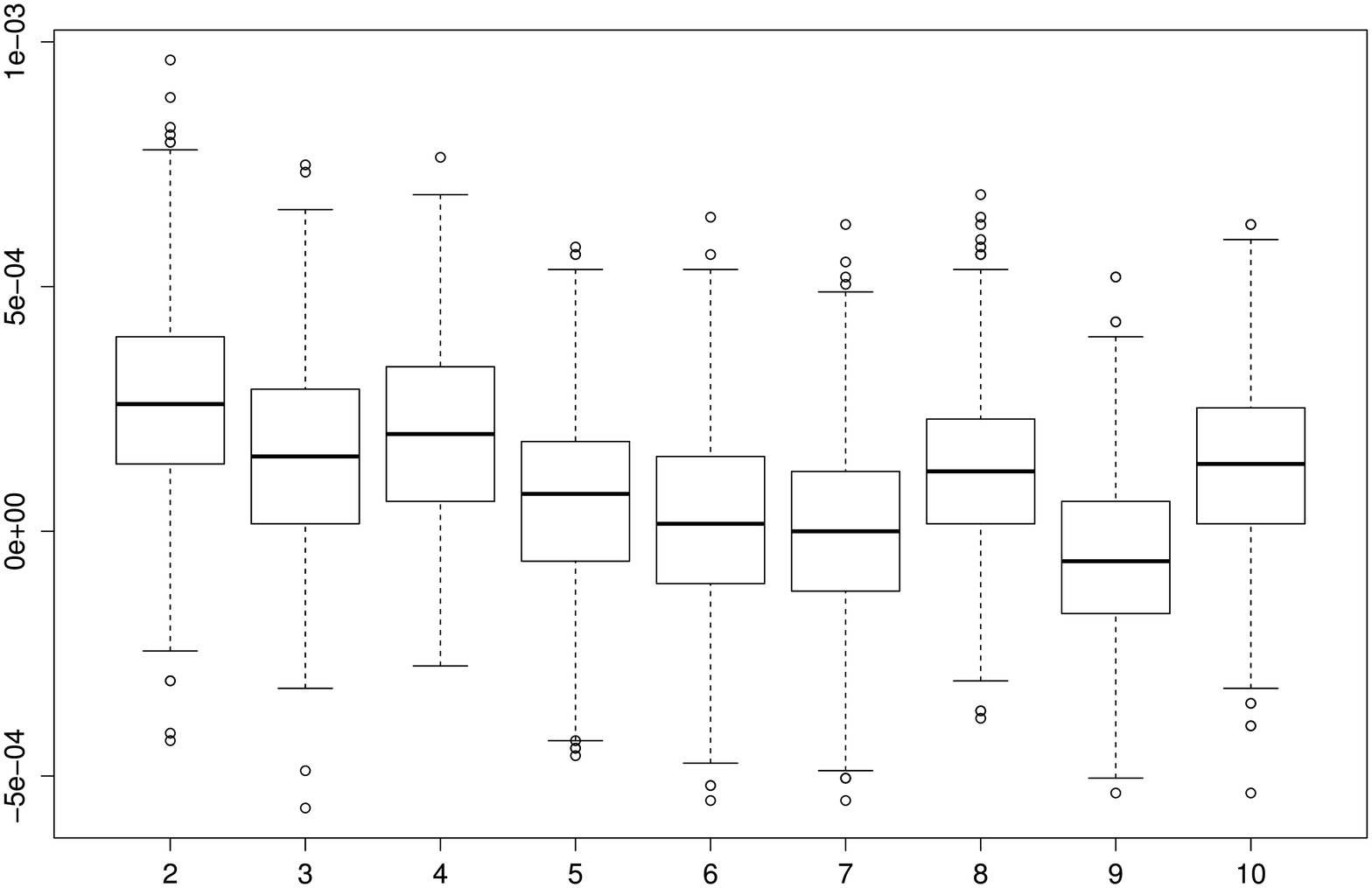}

}
\vspace{1cm}
\hspace*{-1.7cm}\subfloat[S\&P500 Tick]{\includegraphics[scale=0.32]{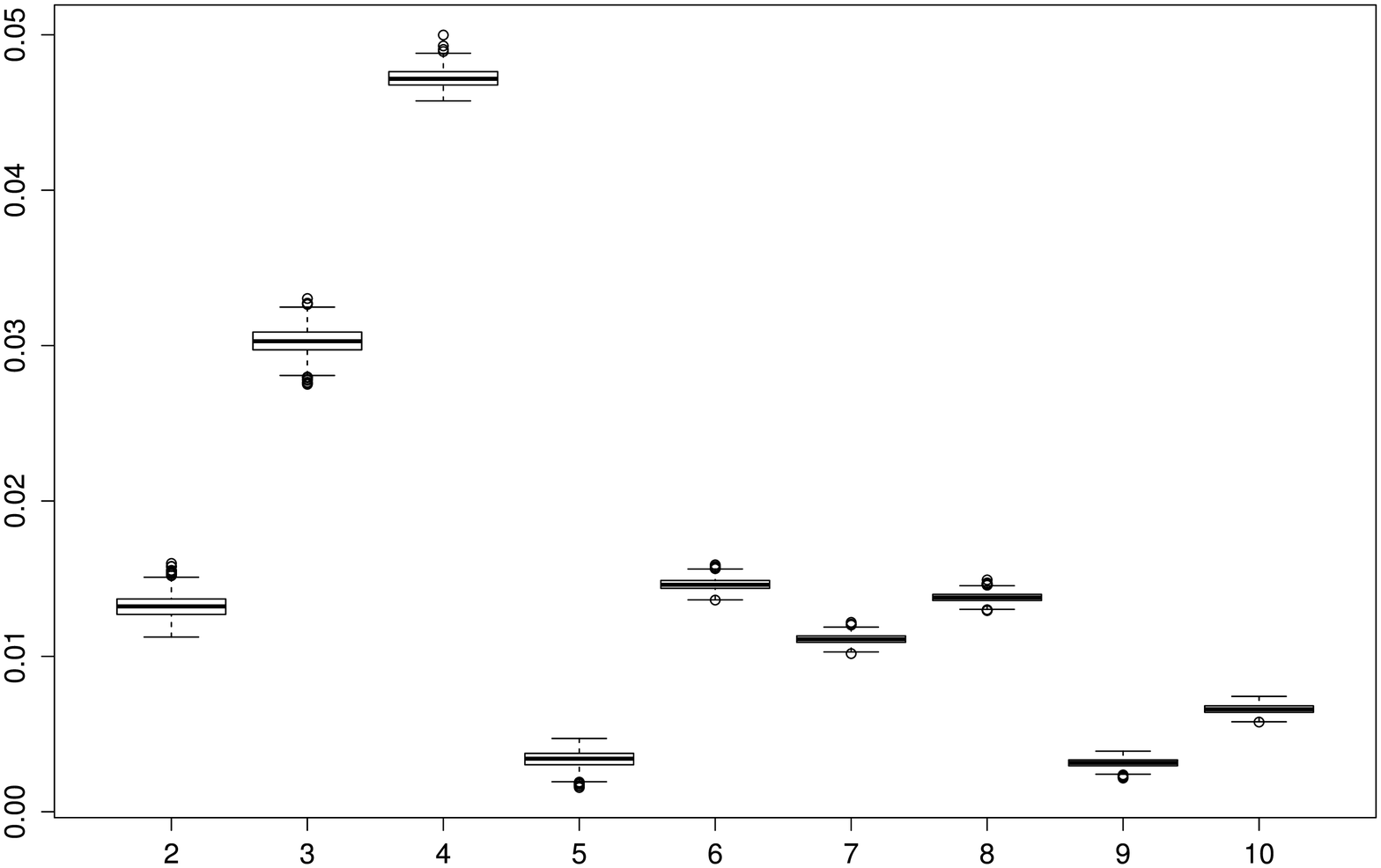}

}\hspace{1cm}\subfloat[PSI Tick]{\includegraphics[scale=0.32]{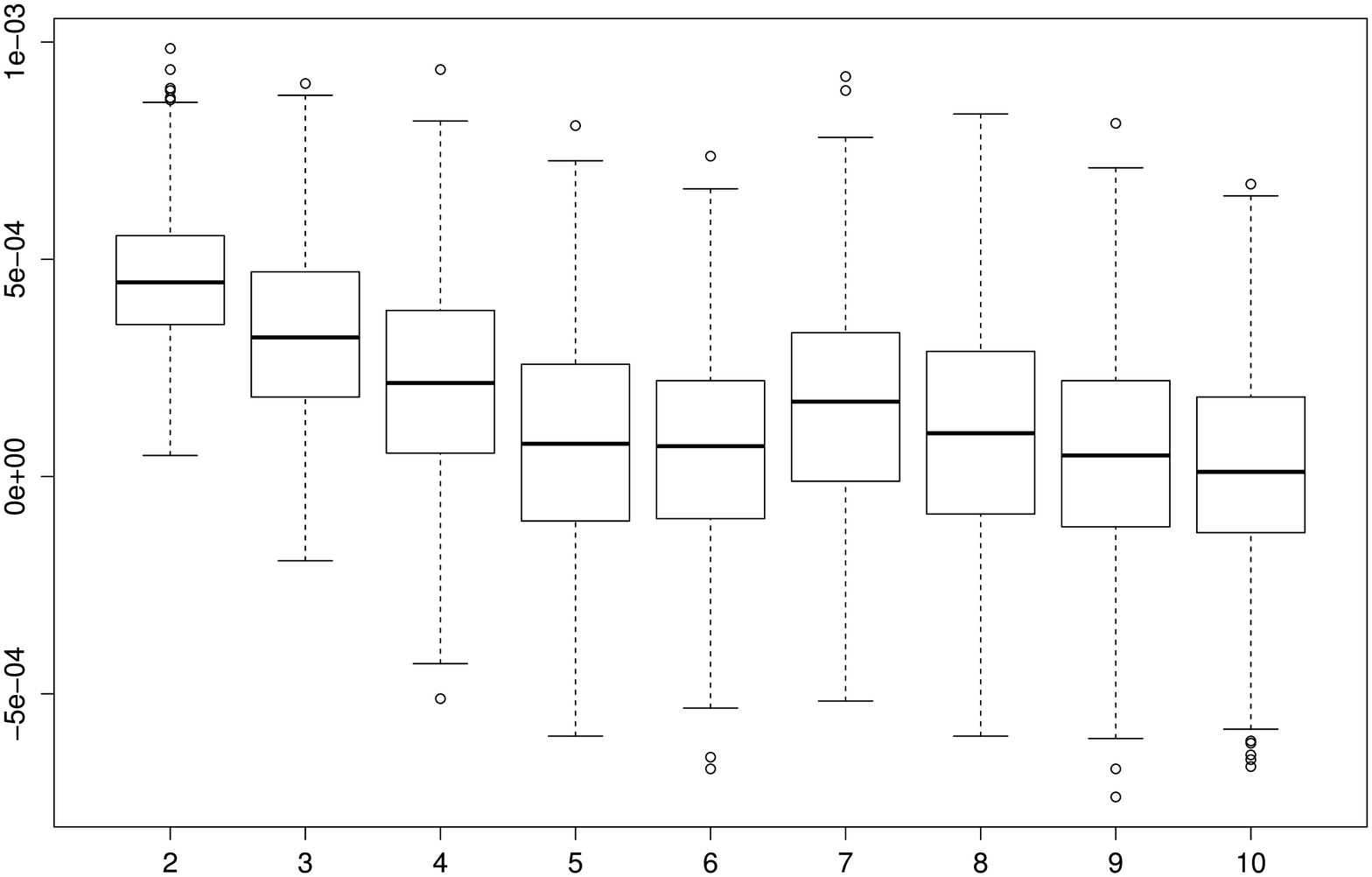}

}

\caption{The serial or lagged dependence function, which is the increase in
the entropy rate estimator at each lag over the entropy rate estimator
at the preceding lag. Confidence intervals are approximate and are
explained in Footnote \ref{fn:neg-cis}. \label{fig:The-serial-dependence-function}}
\end{figure}

In Figure \ref{fig:The-serial-dependence-function}, we plot the increase
in the estimated compression ratio between adjacent block sizes. Block
sizes of lags 2 to 10 in Figure \ref{fig:Compression-ratios} therefore
correspond to lags 1 to 9 in the serial dependence function of Figure
\ref{fig:The-serial-dependence-function}. A point estimate for the
plotted serial dependence function should be considered to be the
mean of the box-and-whisker plots in Figure \ref{fig:The-serial-dependence-function},
while the spreads around the mean shown by the box-and-whisker plots
are an approximation of the (100\% and 75\%) confidence intervals
for this function.%
\footnote{The confidence intervals for the lagged dependence function are approximations
of the true confidence intervals. They are obtained by differencing
the confidence intervals of the compression ratios shown in Figure
\ref{fig:Compression-ratios}. They will be exactly correct if the
distributions of compression ratios are only different in their location
parameter between block sizes, otherwise they will be approximations
of the form $q_{\tau}(\Delta_{k}c_{i})\approx\Delta_{k}q_{\tau}(c_{i})$,
where $q_{\tau}(\cdot)$ is the population quantile function, $\Delta_{k}$
is the first-difference operator between block size $k$ and block
size $k-1$, and $c_{i}$ is the $i^{\mathrm{th}}$ compression ratio
of the shuffled returns series. \label{fn:neg-cis}%
} The lagged serial dependence function with its associated confidence
intervals provides a natural mechanism for identifying the appropriate
lag in an intertemporal model of stock returns and is a generalisation
of the autocorrelation function.

Another measure that would be of interest to study in order to identify
appropriate lags for modelling is the gap between compression ratios
for the shuffled sequences and the compression ratio of the original
squence. Contrary to the serial dependence function, these compression
ratio gaps would be decreasing with increasing block size. Using this
measure, a modeller would be able to find the appropriate lag for
intertemporal modelling by finding the block size whose gap is within
a given tolerance away from 0.

\subsection{A summary of the proposed test for serial dependence\label{sub:A-summary-of}}

For convenience, but at the cost of some redundancy(!), we summarise the steps in the test we propose for serial dependence. The test can be used for arbitrary subsets of random variables, but here we only discuss sets of consecutive returns starting at 1, in the hope that the extension is clear. These tests of dependence among subsets of random variables are highly analogous to likelihood ratio tests of sets of parameters in regression analysis, for example. 

Given some sample data $\{x_{t_1}, ..., x_{t_n}\}$, generated by a stationary ergodic stochastic process $\{X_n\}$, we test the null hypothesis:
\begin{center}
 $H_0:$ for a given $1 < k \leq n$, ${X_1, ..., X_k}$ are jointly independent random variables.
\end{center}
Note that, choosing $k = n$, we can test for independence of $X_1,...,X_n$.
To test the null hypothesis, we proceed as follows:
\begin{enumerate}
\item Compute the empirical distribution function $\hat F(x)$ for the data, as defined in \eqref{eq:cdf}.

\item Map each sample point $x_{t_i}$ to its discretised version\footnote{Here we assume a resolution of $8$ bits, but a different resolution can be used.} using the transformation $\tilde x_{t_i} \longmapsto \lfloor 2^8 \hat F(x_{t_i}) \rfloor$.
         The new collection of points constitutes approximately a sample of the process $\{\tilde X_n =  \lfloor 2^8 F(X_n) \rfloor \}$.
         Note that under the null hypothesis, the random variables $(\tilde X_1, ..., \tilde X_k)$ are independent.
         
         For notational convenience, we denote each block $(\tilde X_{ki + 1}, ..., \tilde X_{k(i+1)})$ by $\tilde Y_i$, for $i = 0,..., I := \lfloor n/k \rfloor -1$.
   
\item Given a random permutation $\sigma$ of $k$ elements, write the sequence $\tilde y_{\sigma(1)}, ... , \tilde y_{\sigma(I)}$ to a binary file and compress it. Given a random permutation $\phi$ of $n$ elements, write the sequence $\tilde x_{t_{\phi(1)}}, \ldots, \tilde x_{t_{\phi(n)}}$ to a binary file and compress it. Store the difference between the first and second compression ratios.
Repeat this step several times\footnote{In our simulations we repeated this step $1000$ times.} to obtain a distribution of the difference in compression ratios under sampling variation. Let the empirical quantile function for this distribution be denoted $Q(\cdot )$.
Note that, under the null hypothesis, the distributions of $(\tilde Y_{\sigma(1)}, ... , \tilde Y_{\sigma(I)})$ and $(\tilde X_{t_{\phi(1)}}, \ldots, \tilde X_{t_{\phi(n)}})$ are the same. Therefore, under the null hypothesis, the distribution computed before should be concentrated around $0$.
\item Given a size for the test $\alpha\in[0,1]$, reject the null hypothesis when $0<Q(\alpha)$.
\end{enumerate}

%

We can quickly exemplify the application of this test to the daily,
minute, and tick return data of the S\&P and PSI. Indeed, for $k$=
2, the distribution of the difference of compression ratios computed
in step 3 is represented in the left-most point of the box-whisker
plots in Figure \ref{fig:The-serial-dependence-function}. A quick
analysis of these plots reveals that the independence hypothesis for
consecutive returns would be rejected at all confidence levels for
the tick returns of the S\&P and the PSI, and for the minute returns
of the S\&P. Furthermore, the same hypothesis would be rejected at
most confidence levels for the minute returns of the PSI and daily
returns of PSI and S\&P. In other words, for most confidence levels,
our test rejects the hypothesis of independence between consecutive
returns of the S\&P and PSI for daily, minute and tick frequencies.

\subsection{Further examples of identifying lagged dependence \label{sub:Further-examples}}

In this section, we introduce several other examples. Some of these
examples come from synthetic data and serve to corroborate the power
and robustness of our test. We also include an example, where our
methodology is used to assess the goodness of fit of a GARCH(1,1)
model, by studying the intertemporal structure of its residuals.

\subsubsection{Separate sample intervals for the S\&P500 returns}

\begin{figure}
\vspace{-2cm}%
\begin{minipage}[t]{1\columnwidth}%
\subfloat[Daily1 entropy rate estimators]{\includegraphics[scale=0.25]{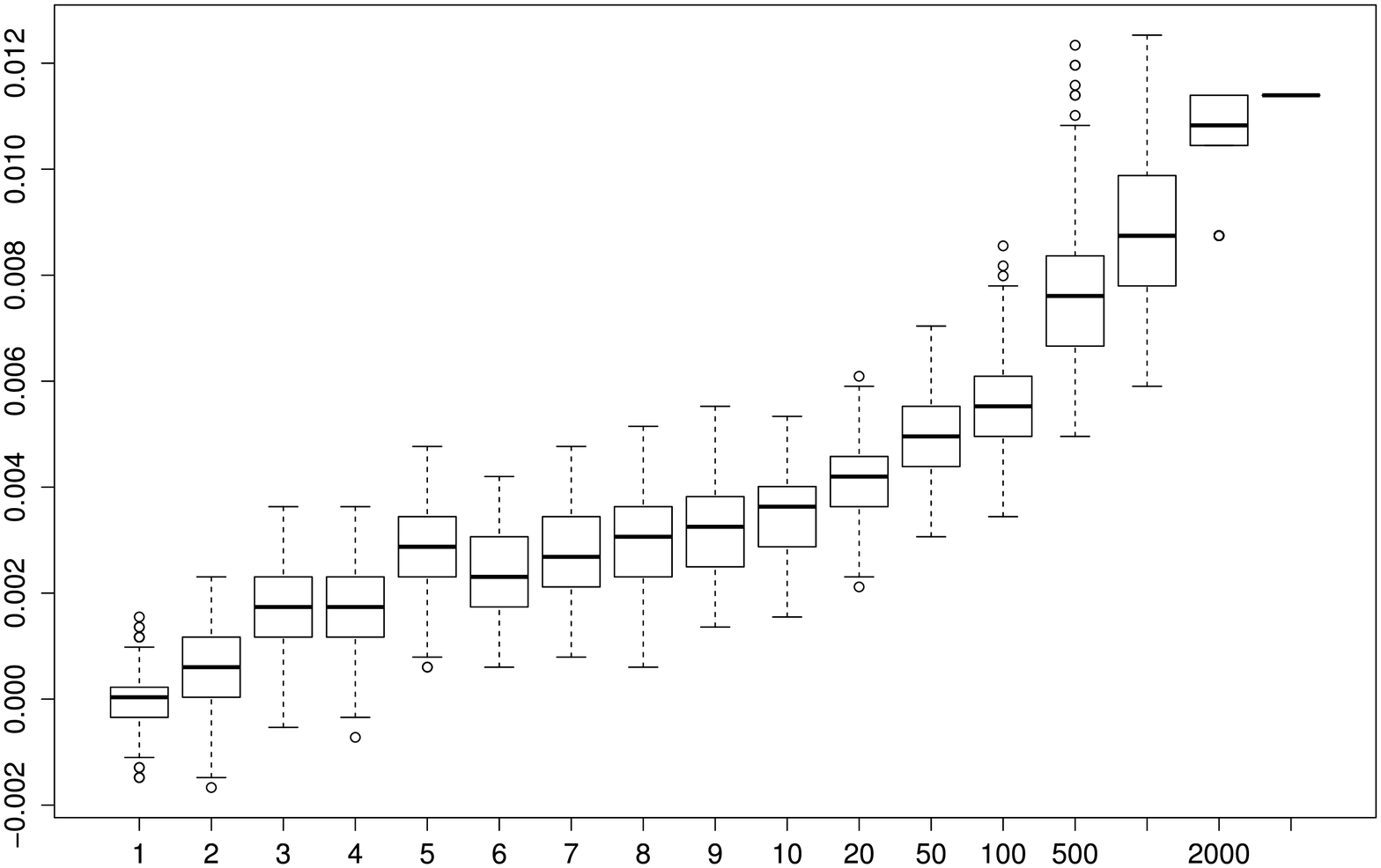}

}\hspace{1cm}\subfloat[Daily2 entropy rate estimators]{\includegraphics[scale=0.25]{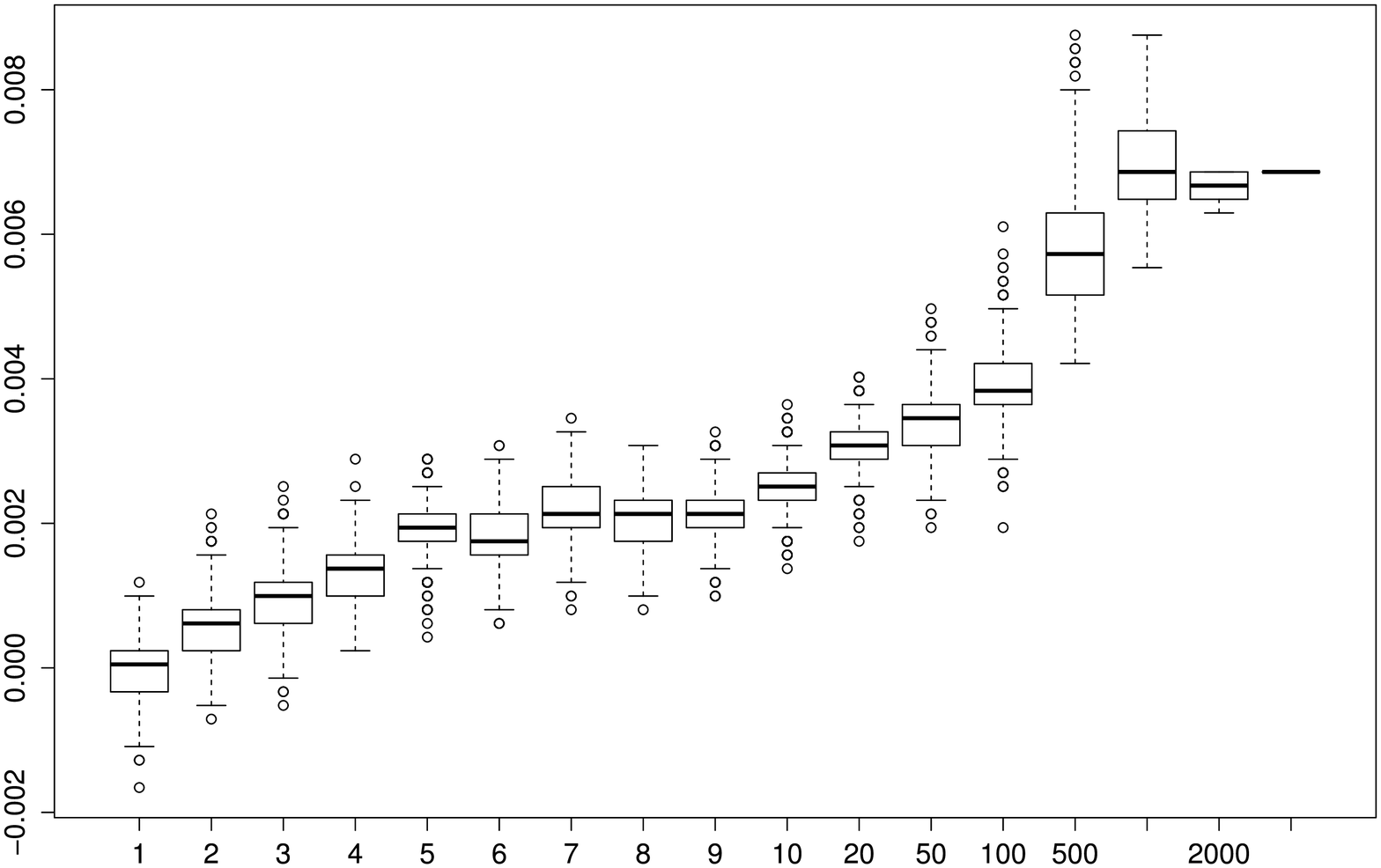}

}
\vspace{0.7cm}
\subfloat[Daily1 lagged dependence function]{\includegraphics[scale=0.25]{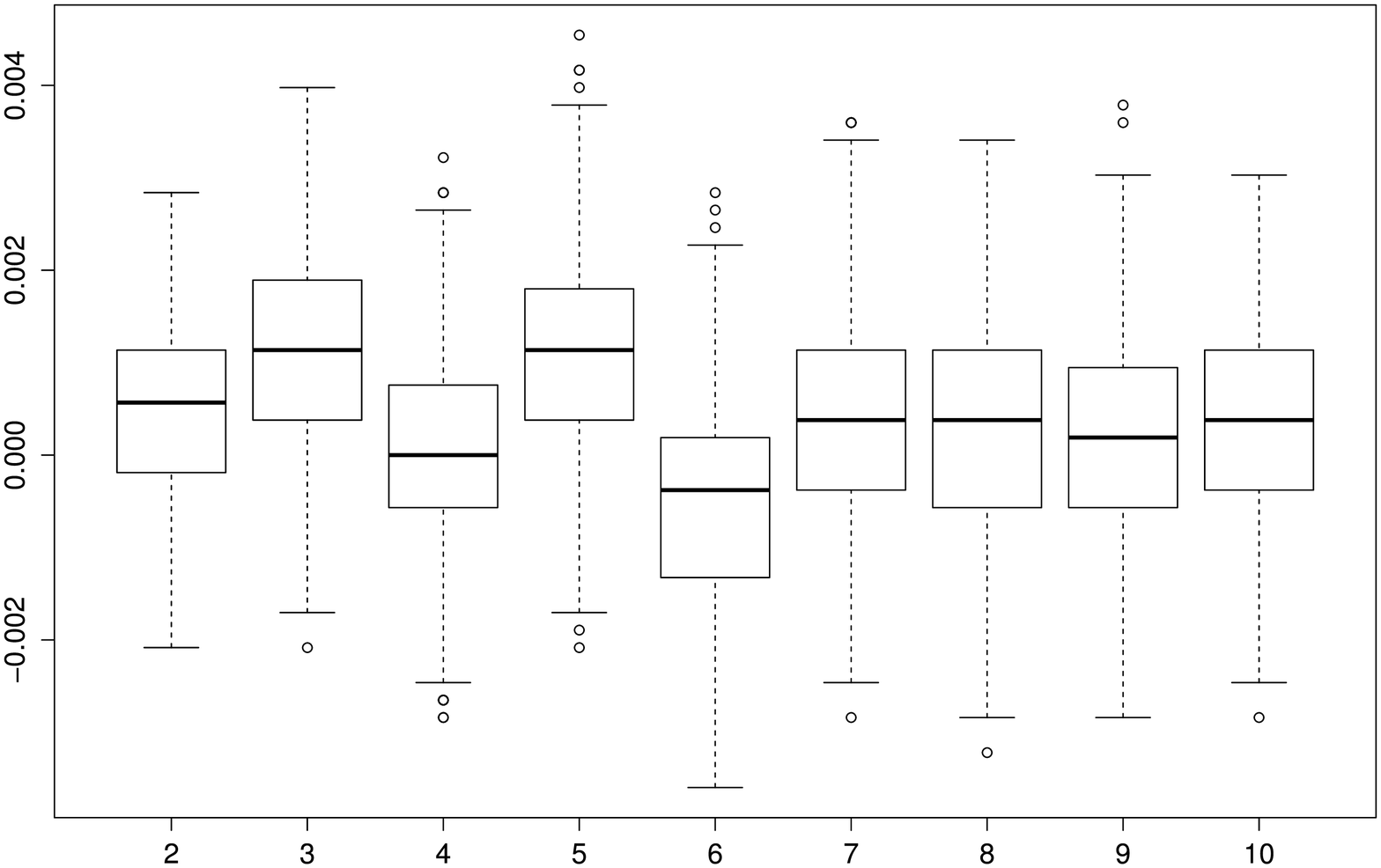}

}\hspace{1cm}\subfloat[Daily2 lagged dependence function]{\includegraphics[scale=0.25]{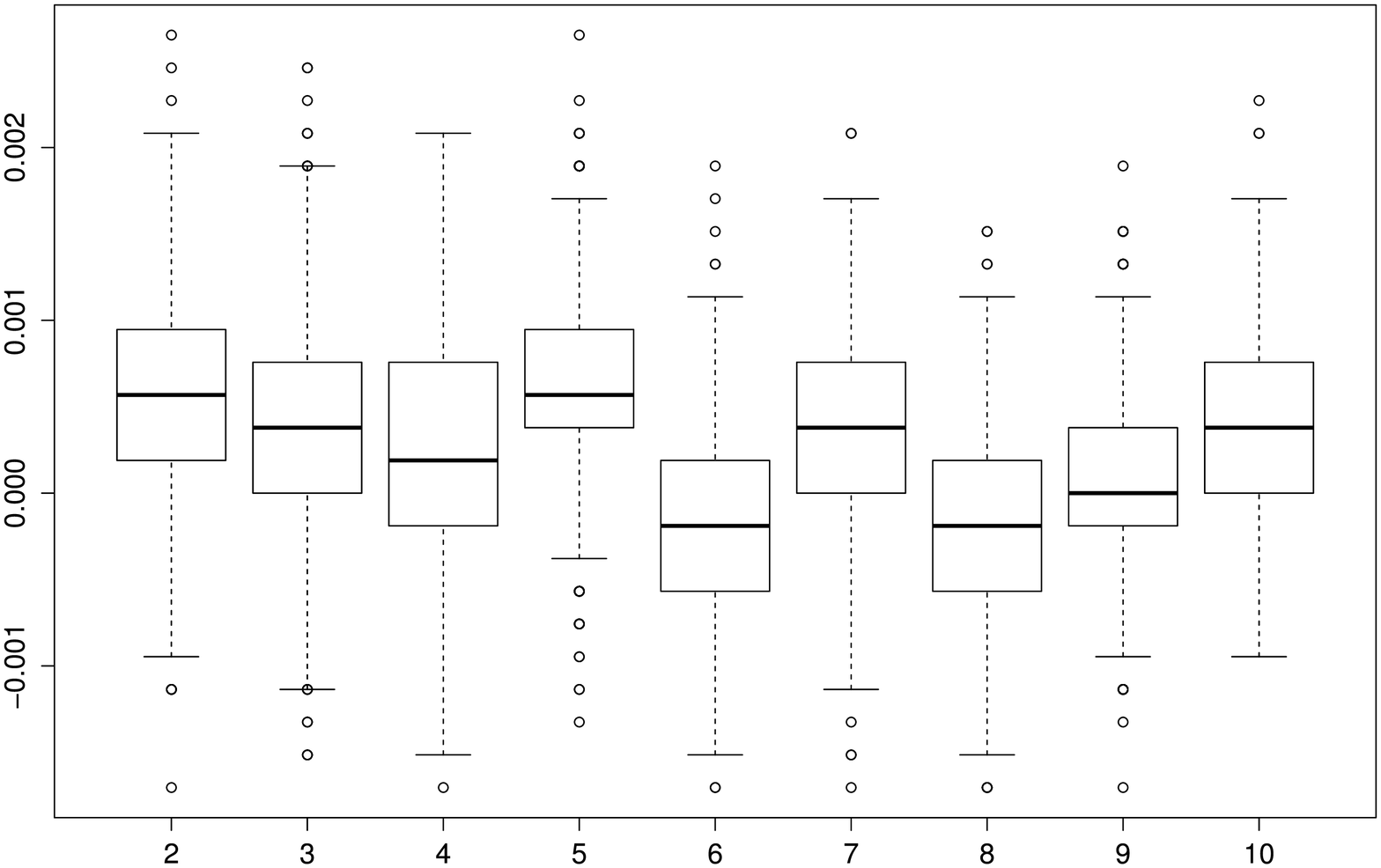}

}
\caption{Comparing the early and late sample periods of the daily S\&P500 returns
series, we can see that the compression rate estimate is greater in
the earlier sample. \label{fig:sp-early-late}}
\end{minipage}

\begin{minipage}[t]{1\columnwidth}%
\vspace{0.7cm}
\subfloat[10,000]{\includegraphics[scale=0.25]{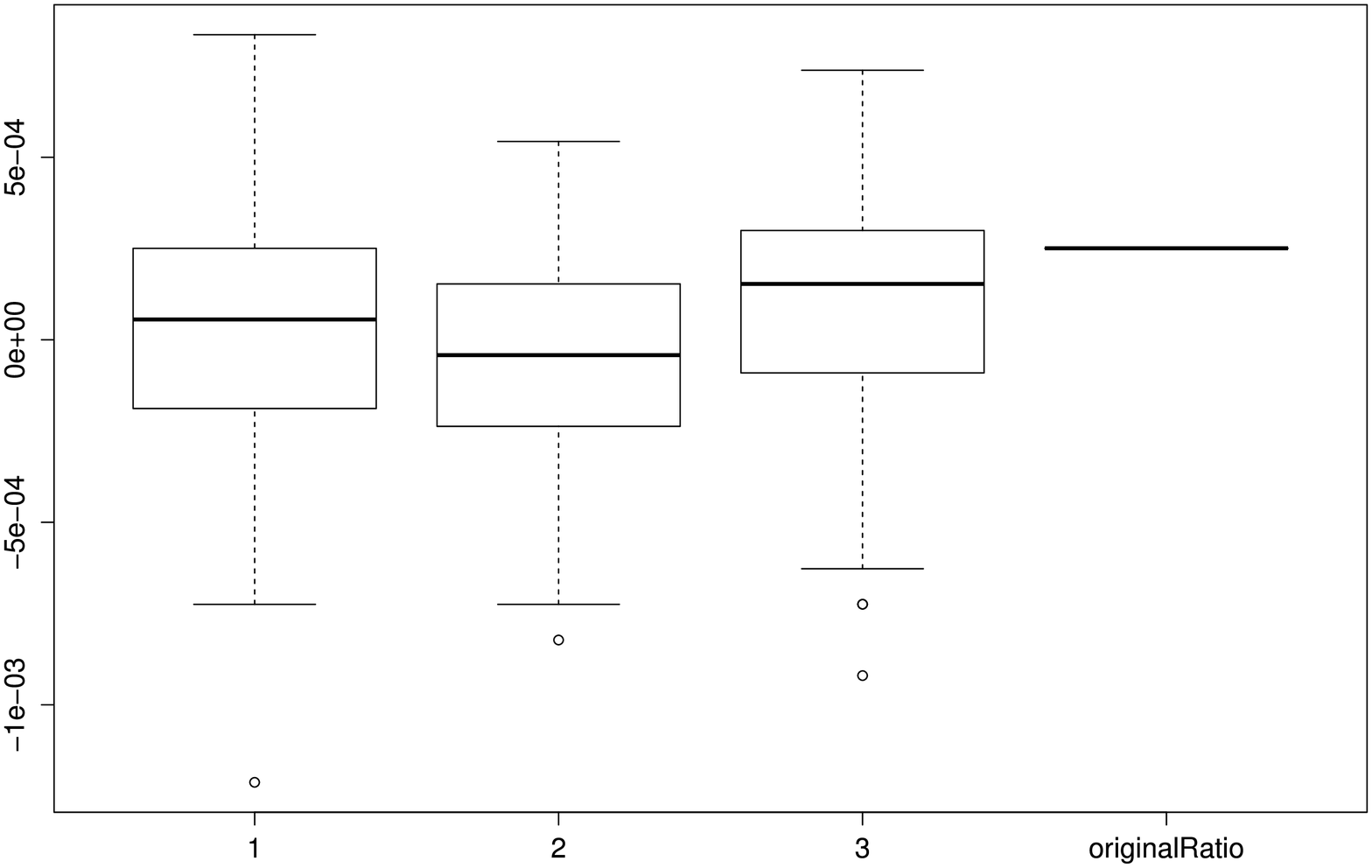}

}\hspace{1cm} \subfloat[50,000]{\includegraphics[scale=0.25]{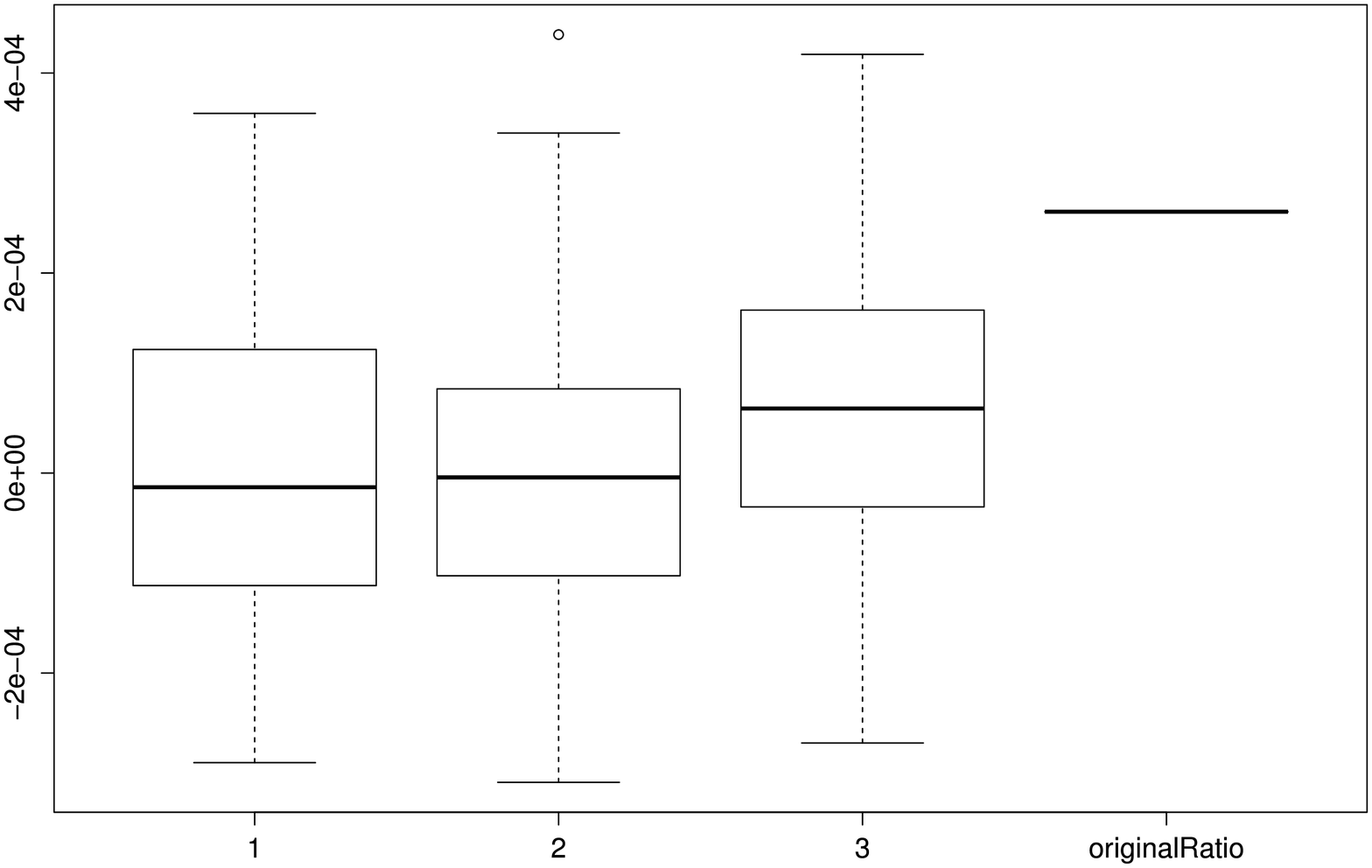}

}
\vspace{0.7cm}
\subfloat[100,000]{\includegraphics[scale=0.25]{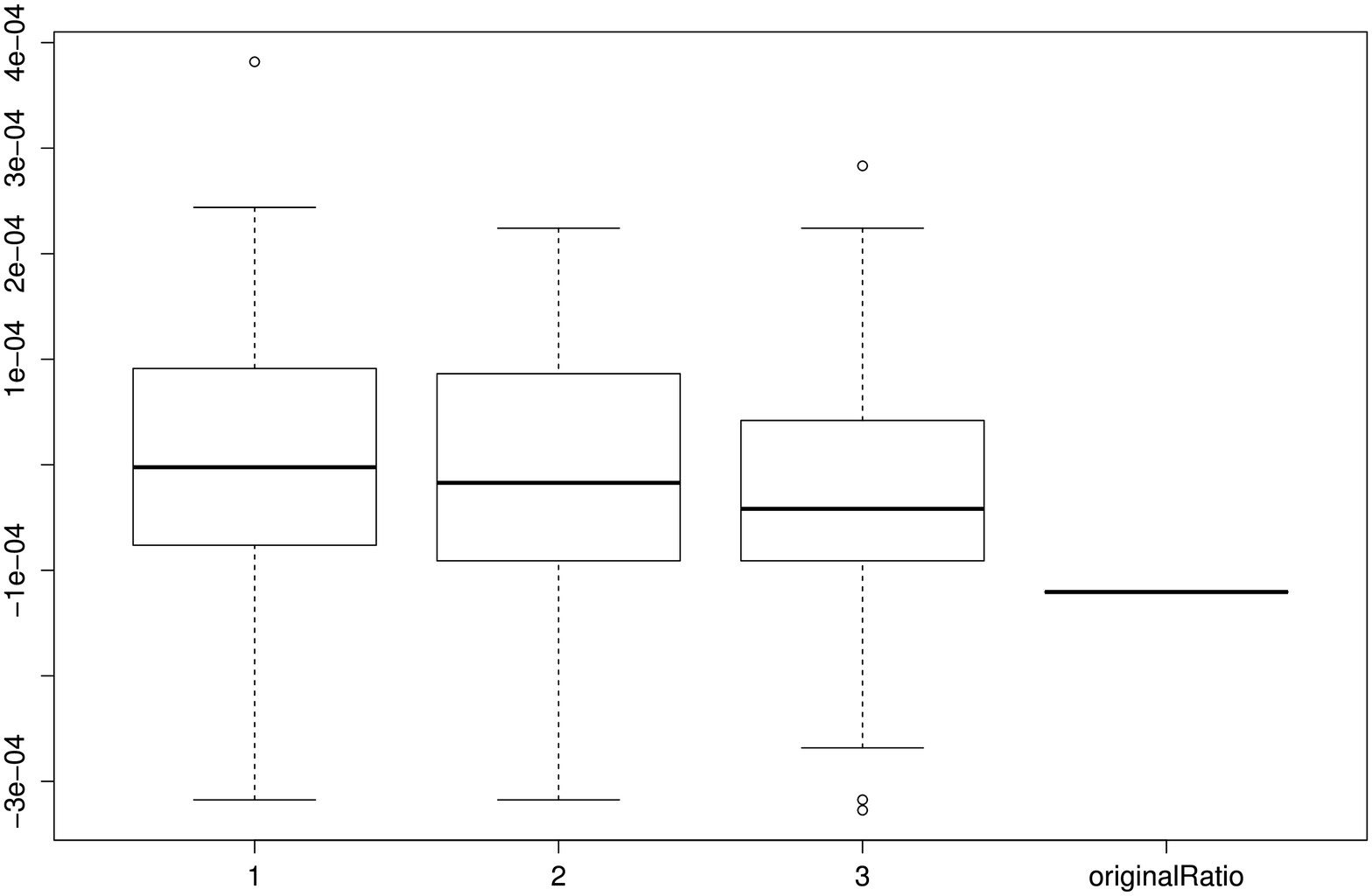}

}\hspace{1cm} \subfloat[500,000]{\includegraphics[scale=0.25]{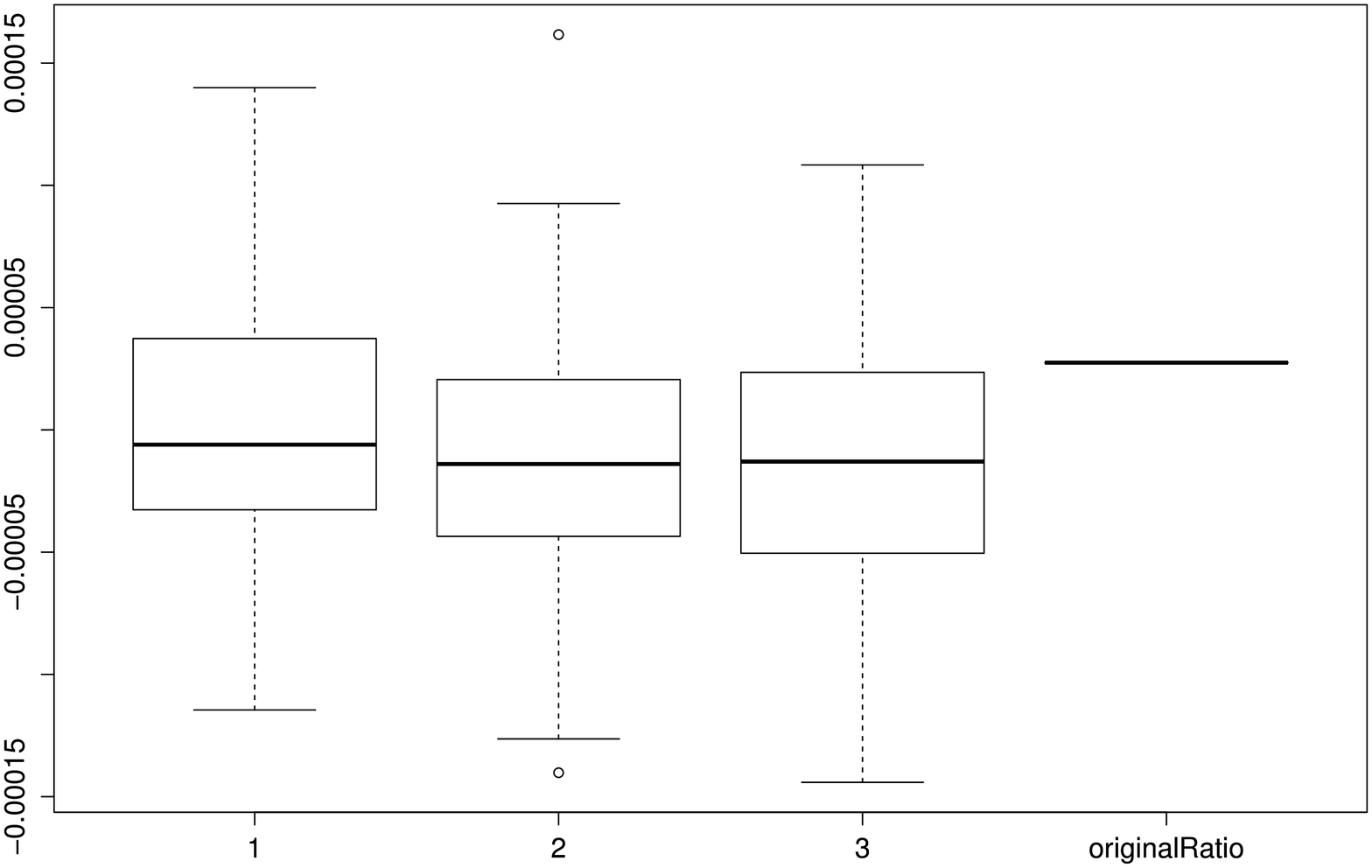}

}

\caption{Compression rate estimators against block size for the Brownian motion
model of log stock prices. \label{fig:rw}}
\end{minipage}
\end{figure}

A comparison of the tests for serial dependence between the early
and late samples of the S\&P500%
\footnote{Summary statistics for the returns in the two periods can be found
in Table \ref{tab:Summary-statistics}.%
}, is presented in Figure \ref{fig:sp-early-late}. The plots show
that S\&P500 returns showed much more intertemporal structure in the
early sample than in the late sample as measured by the entropy rate.
The compression ratio in the early sample is estimated at about $1.1\%$,
while that of the late sample is about $.7\%$. Furthermore, at minimum,
about 1000 lags are required to describe the behaviour of the returns
in the early sample, while about 500 may be acceptable in the late
sample. 

\subsubsection{Brownian motion model for log prices}

If log prices are described by a Brownian motion, then returns sampled
on a uniform grid are independent and identically distributed random
variables with normal distribution. We simulated one realisation of
a random walk over 500,000 steps, and we show the sampling uncertainty
in our estimator for various sample lengths in Figure \ref{fig:rw}.
The estimates for the compression rate in the first 10, 50, 100 and
500 thousand steps are all close to zero, which is the correct quantity.
The confidence intervals obtained from our shuffling procedure reliably
cover the point estimator.

\subsubsection{Hidden dependence in time series}

\begin{figure}
\subfloat[10,000]{\includegraphics[scale=0.25]{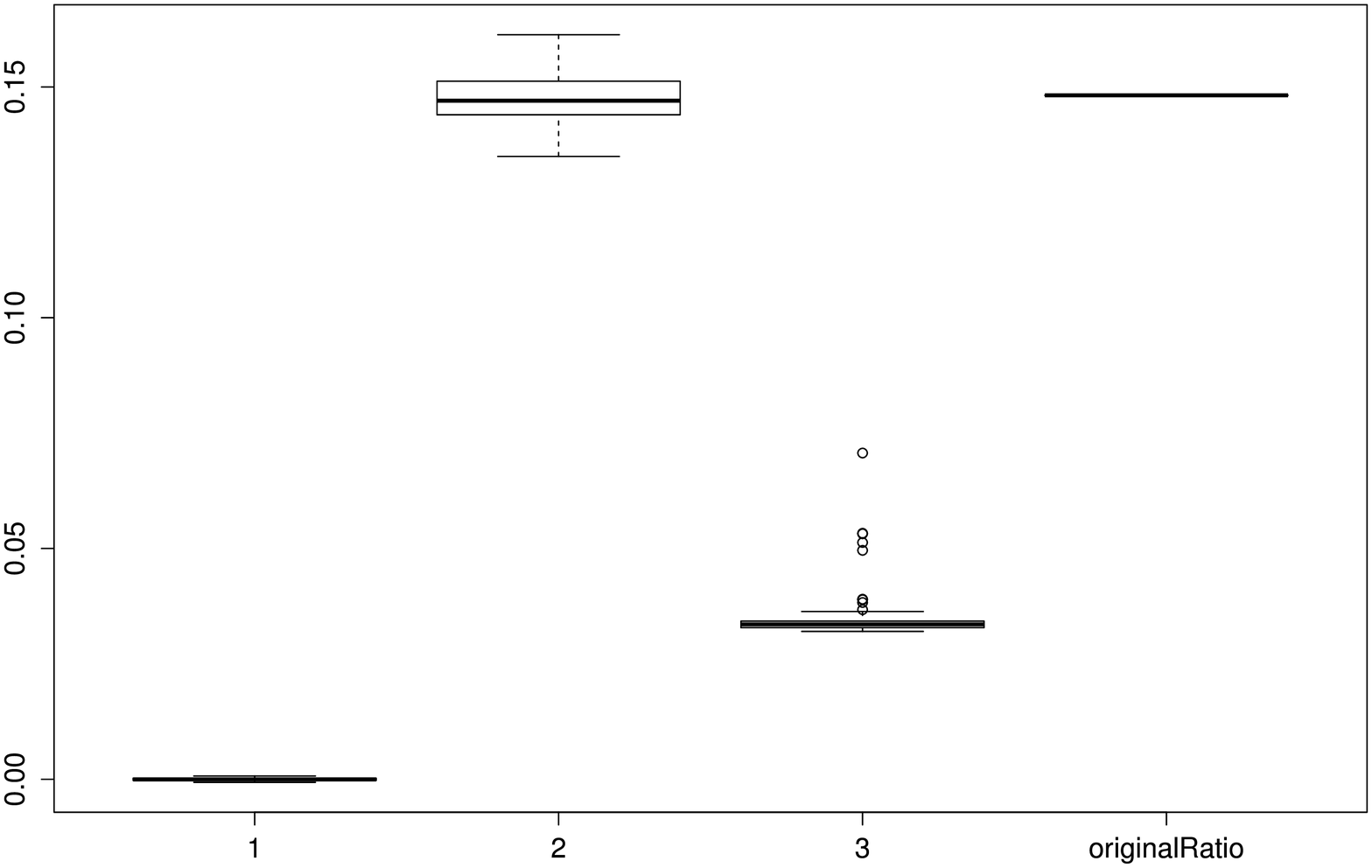}

}\hspace{1cm} \subfloat[50,000]{\includegraphics[scale=0.25]{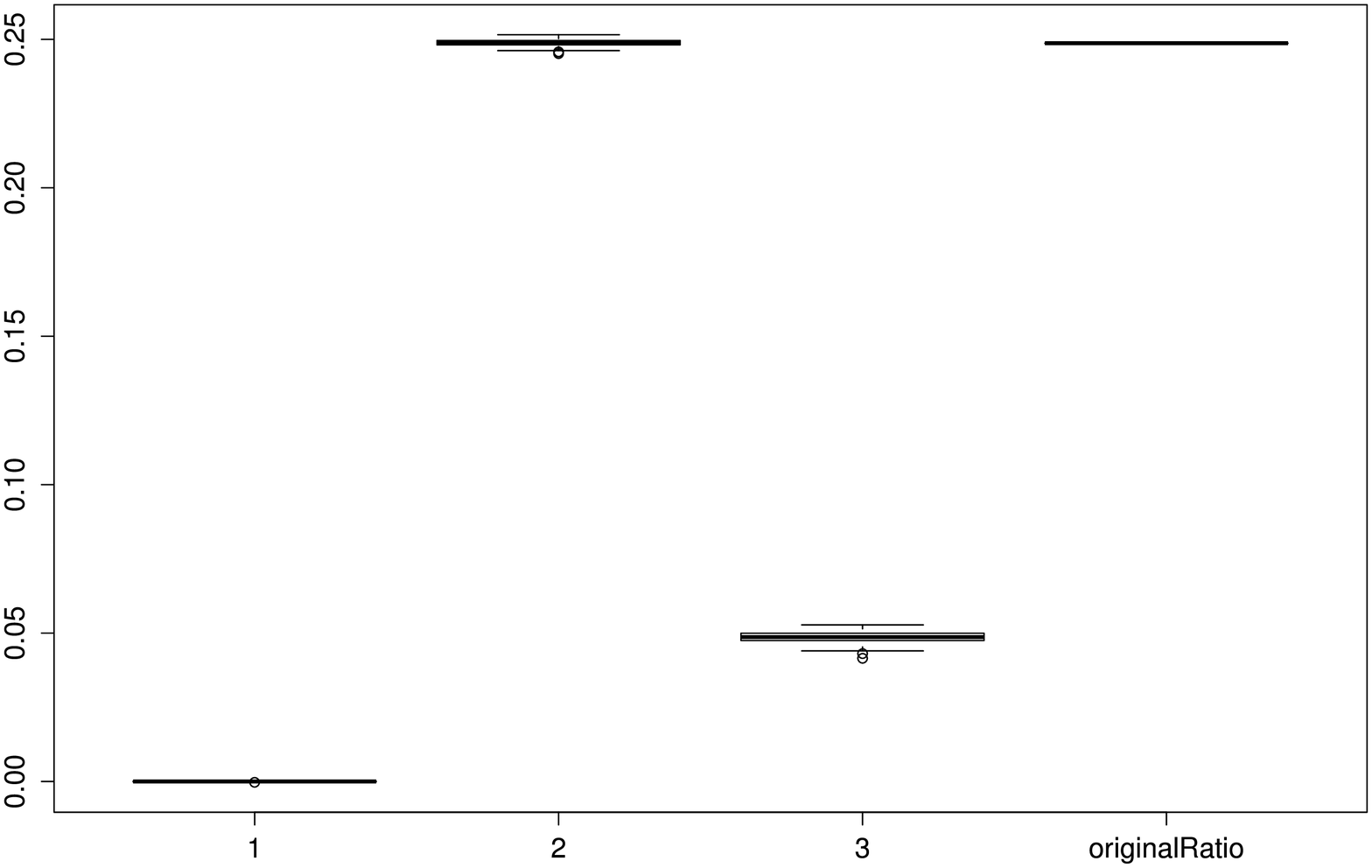}

}
\vspace{1cm}

\subfloat[100,000]{\includegraphics[scale=0.25]{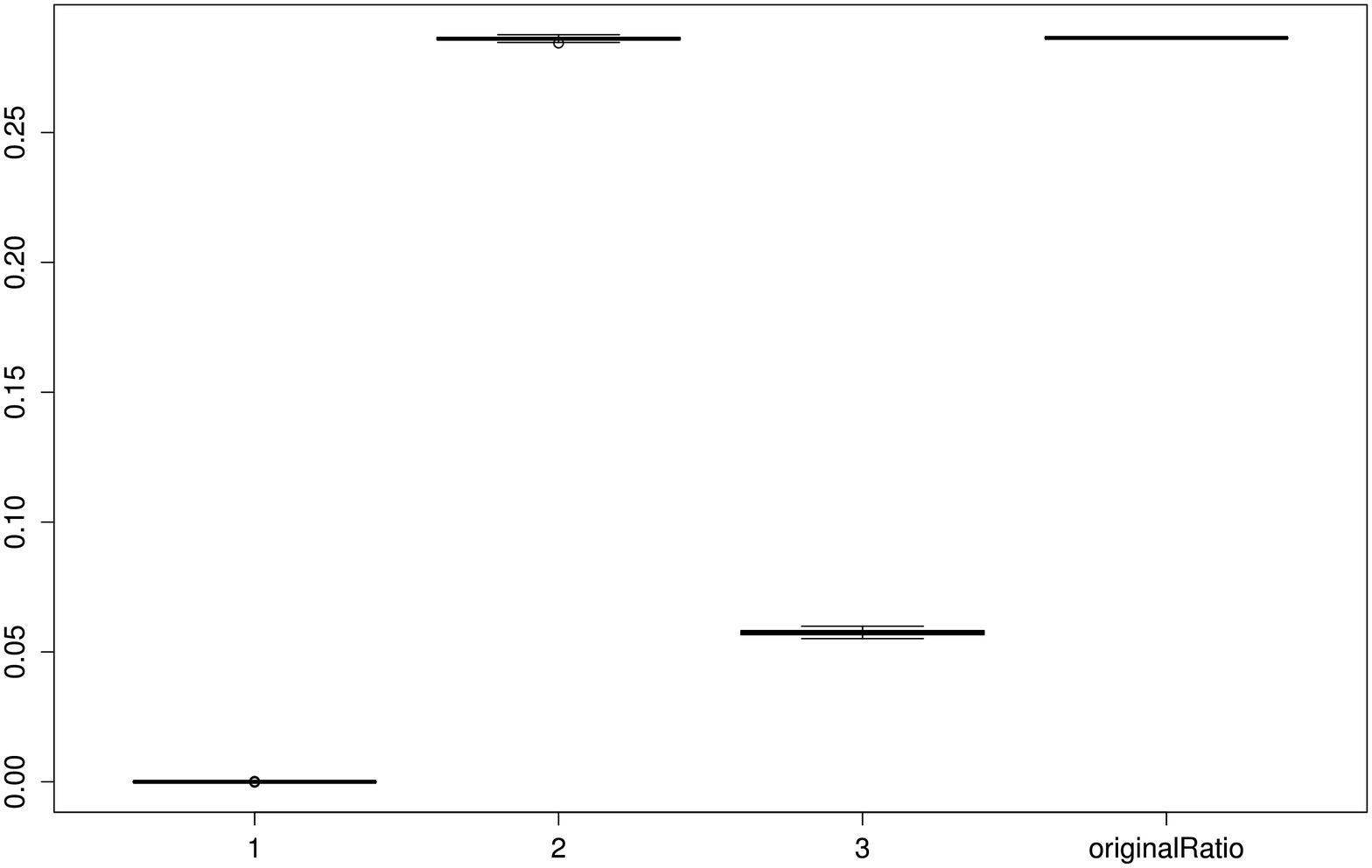}

}\hspace{1cm} \subfloat[500,000]{\includegraphics[scale=0.25]{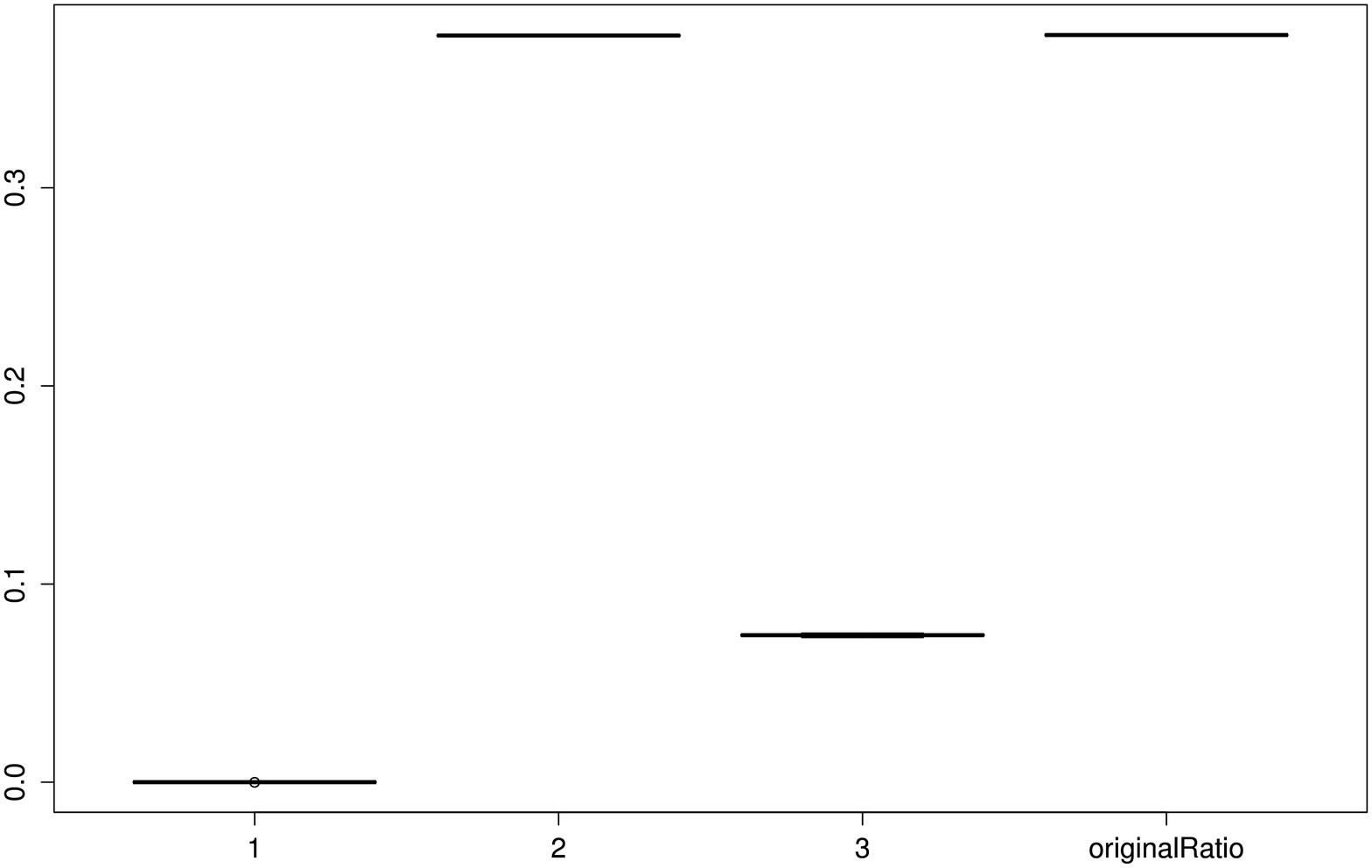}

} \vspace{1cm}

\subfloat[1,000,000]{\includegraphics[scale=0.25]{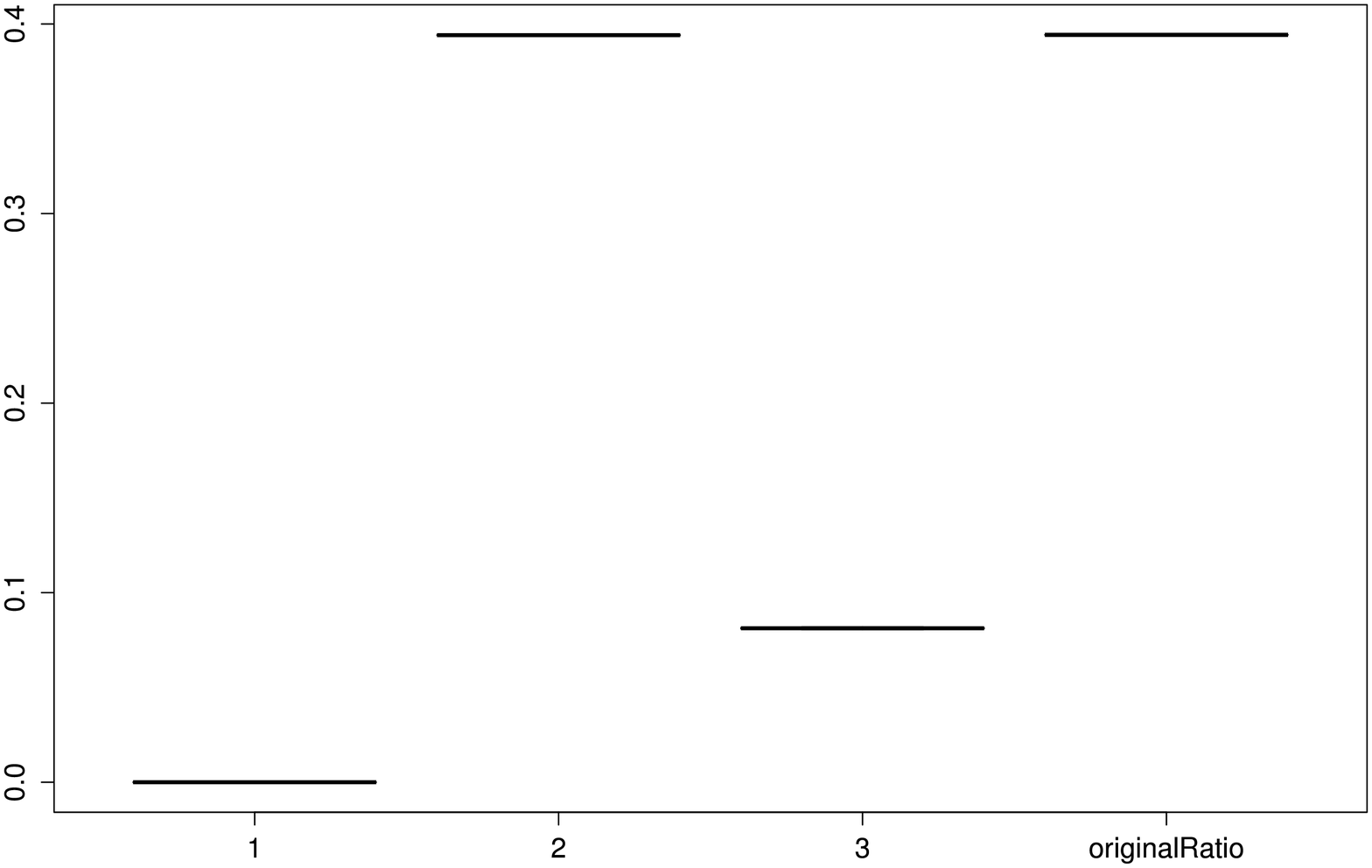}

}\hspace{1cm} \subfloat[ACF 1,000,000]{\includegraphics[scale=0.25]{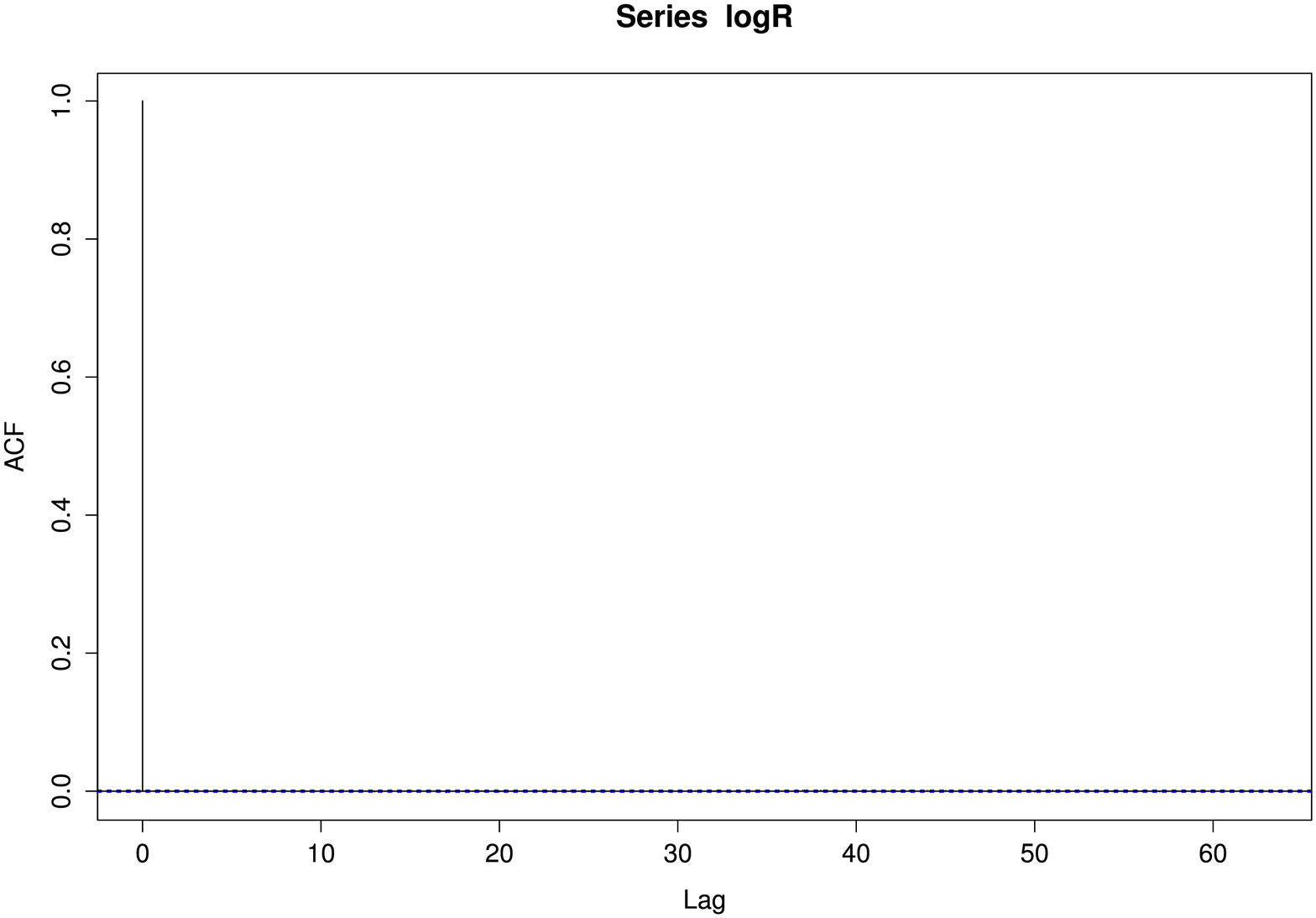}

}

\caption{Subfigures (a)-(e) show the compression rate estimates for the hidden
dependence model at sample lengths 10, 50, 100, 500 and 1000 thousand
observations. Subfigure (f) shows the autocorrelation function of
the sample with length 1,000,000. \label{fig:hidden}}
\end{figure}

We construct an artificial example of a particularly challenging stochastic
process to model, and we use our method to uncover its intertemporal
dependence structure. Consider the stochastic process $\{X_{n}\}$
defined on the positive integers $\mathbb{N}$ by 
\[
X_{n}=\begin{cases}
\epsilon_{i} & n=2i-1\,\,\,\exists i\in\mathbb{N}\\
|\epsilon_{i}|\left(2\mathbb{I}\{|\epsilon_{i}|>1.19\}-1\right) & n=2i\,\,\,\exists i\in\mathbb{N}
\end{cases}
\]
for $\epsilon_{i}\overset{iid}{{\sim}}N(0,1)\,\,\,\forall i\in\mathbb{N}$,
so that the pairs $(X_{2i-1},X_{2i})$ are fully dependent, while
any other pairs are not. The number $1.19$, was chosen so that the
autocorrelation of the process would be approximately $0$ for all
lags greater than $1$. In this case, an analysis based on use of
the autocorrelation function would erroneously suggest independence.
Nevertheless, our method is able to discover its dependencies. We
simulate one realisation of one million observations from this process
and plot the results for increasing subsamples in Figure \ref{fig:hidden}.
Even for the first 10,000 observations our method is able to detect
significant intertemporal structure through an compression rate of
$15\%$ at a $p$-value of zero. The drop in dependence at a lag length
of 3 occurs because about half the odd-even pairs are broken in the
shuffling of size 3 blocks, so the method is able to identify the
unique pairwise dependence structure in this example. It is also interesting
to notice how the compression rate approaches the optimal rate of
$50\%$ as the sample size increases.

\subsubsection{GARCH(1,1)}

\begin{figure}
\subfloat[GARCH(1,1) residuals]{\includegraphics[scale=0.25]{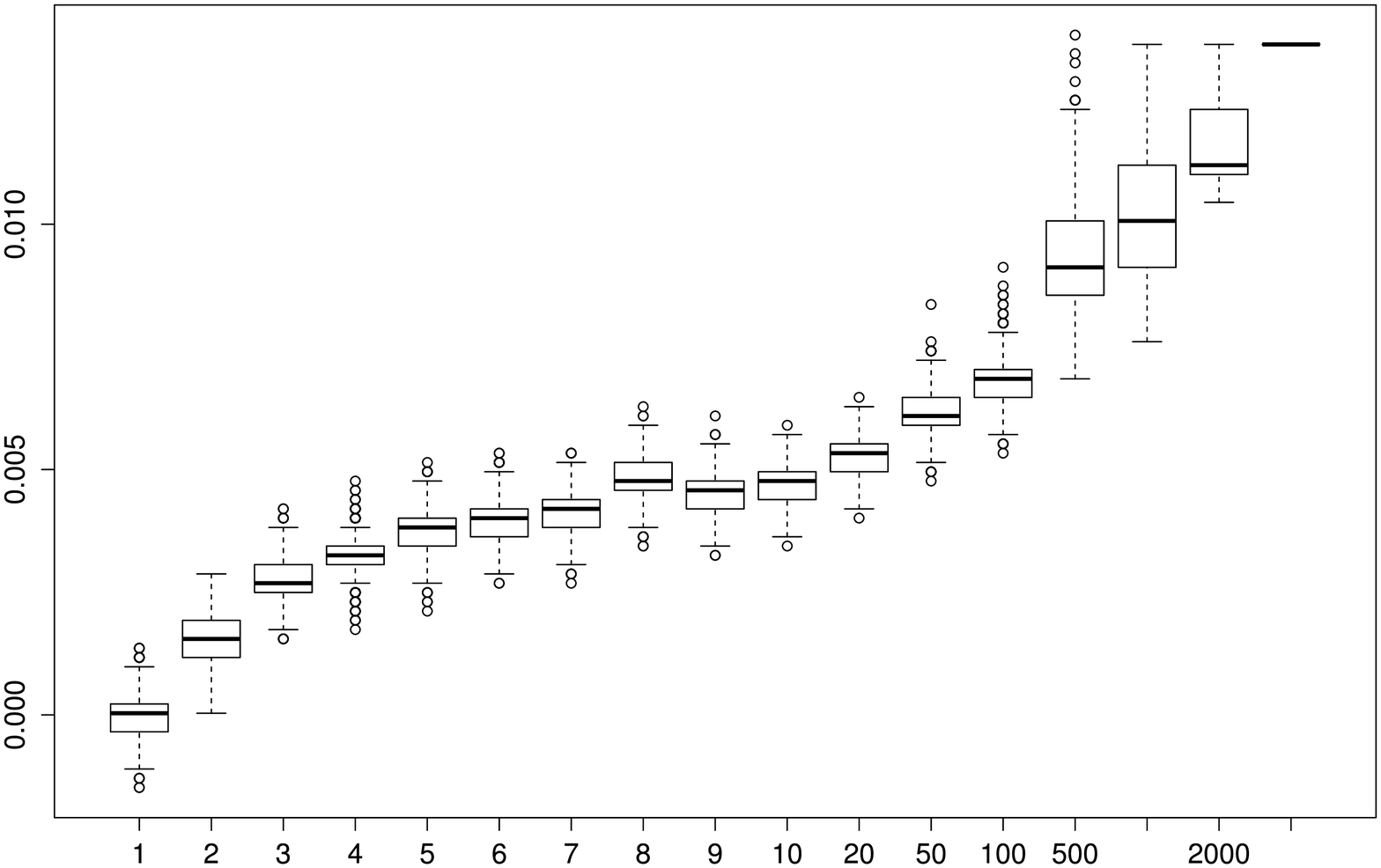}

} \hspace{1cm} \subfloat[GARCH(1,1) simulation]{\includegraphics[scale=0.25]{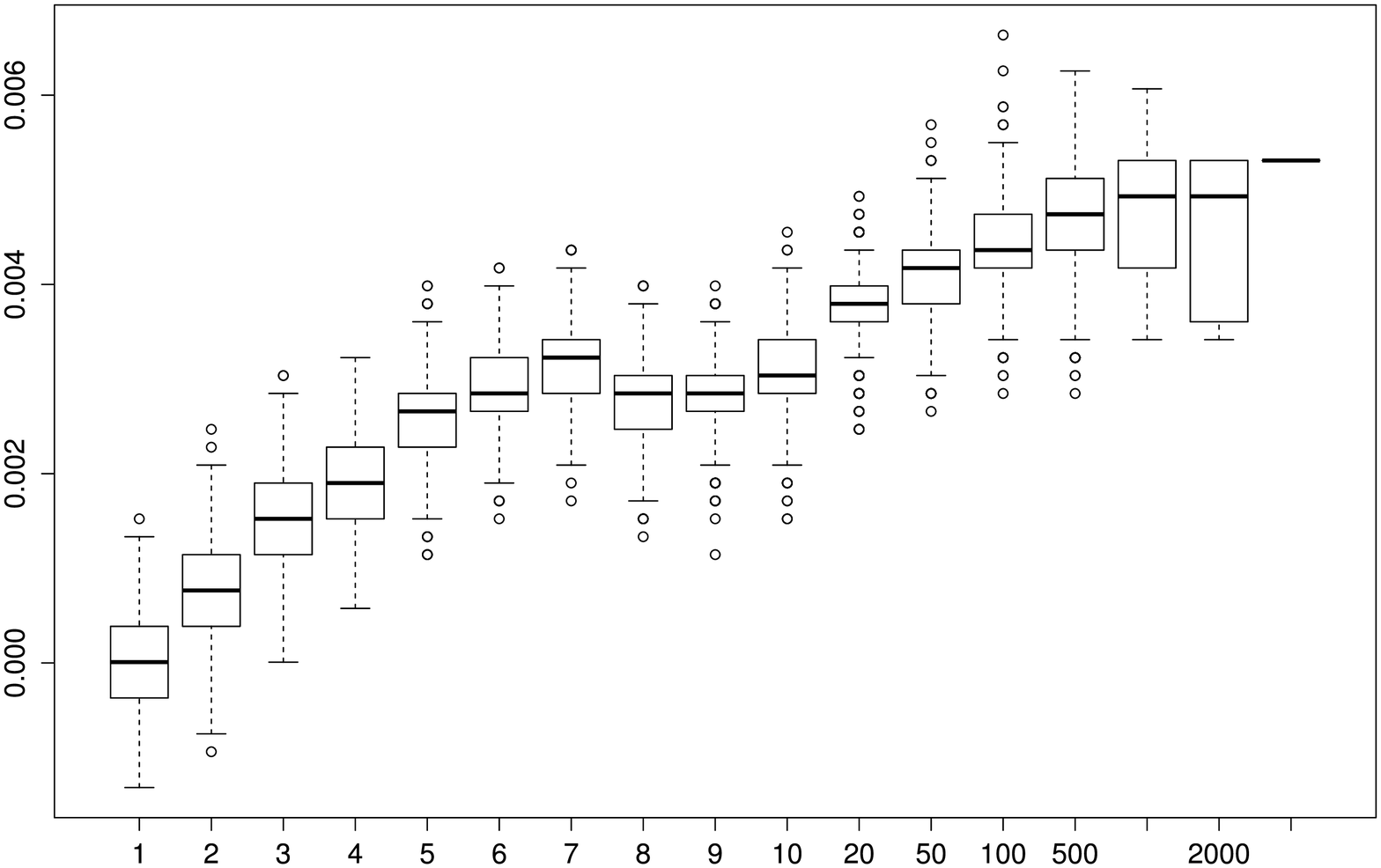}

} \vspace{1cm}

\subfloat[GARCH(1,1) residuals]{\includegraphics[scale=0.25]{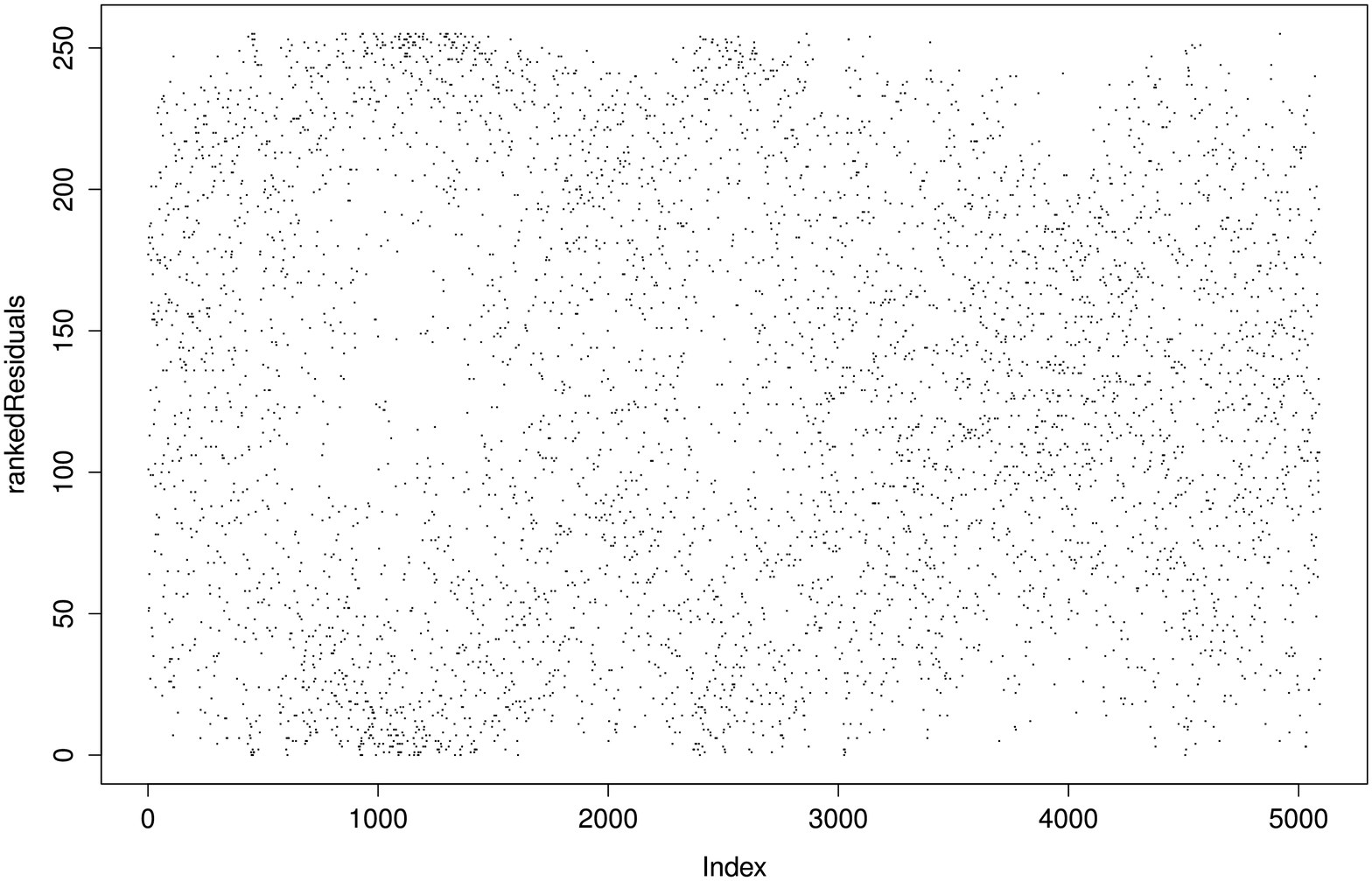}

} \hspace{1cm} \subfloat[GARCH(1,1) simulation]{\includegraphics[scale=0.25]{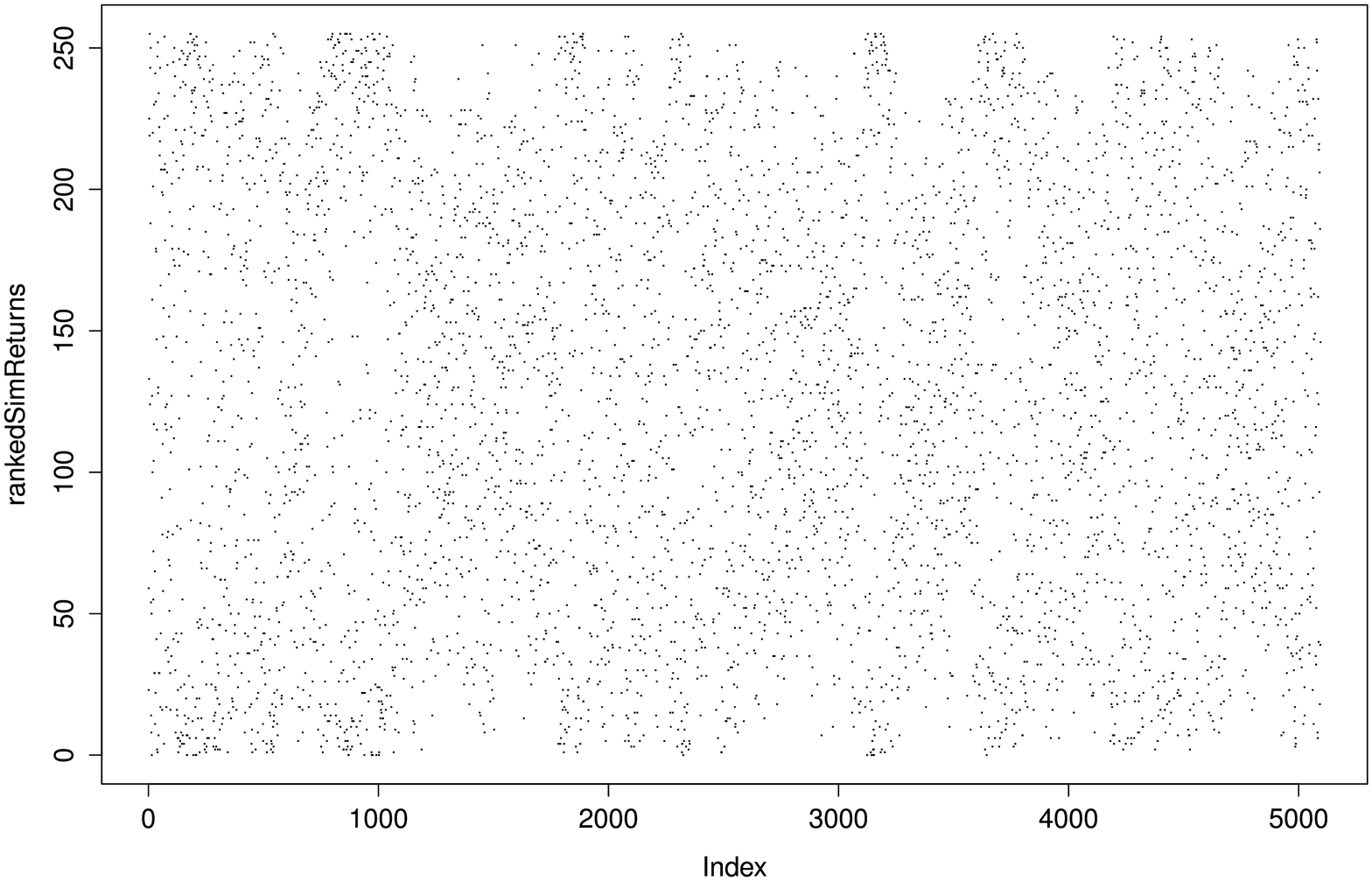}

}

\caption{We fit a GARCH(1,1) model to the Daily1 sample of the S\&P500 returns
series by maximum likelihood. Subfigures (a) and (c) pertain to the
residuals of this model and Subfigures (b) and (d) pertain to a single
random simulation of the fitted model. (a) and (b) are the usual entropy
rate estimator plots, while (c) and (d) are the rank plots at 8 bit
resolution as defined in Section \ref{sub:Visualisation}. \label{fig:garch}}
\end{figure}

We fit a GARCH(1,1) model to the early sample (Daily1 sample) of the
S\&P500 returns series by maximum likelihood. In Figure \ref{fig:garch}
we see that the residuals of this model exhibit substantial dependence
at all but the longest lags. By comparing this plot with Figure \ref{fig:sp-early-late},
we see that the compression rate estimates of the original returns
series and the model residuals are both close to $1.1\%$. The GARCH(1,1)
model therefore does not seem to remove much of the structure in returns,
and the model may in fact introduce more structure than it removes.
By simulating one realisation of the fitted model, we are able to
recover much of its lag structure from the entropy rate estimators
in panel (b). We can also see that the simulated data does not exhibit
as long a memory as the original S\&P500.

\section{Conclusion \label{sec:Conclusion}}

Information theory offers powerful methods that have wide applicability
to the study of economic problems. Using a data compression algorithm
we are able to discover substantial intertemporal dependence in stock
returns in different countries, at different sampling frequencies
and in different periods. We argue that this dependence does not amount
to evidence against the efficient markets hypothesis, in contrast
to some earlier research. We make the observation that statistical
redundancy is qualitatively equivalent to evidence of dependence,
and we show how to test for intertemporal dependence in time series
using an estimator for entropy rate based on a data compression algorithm.
By proposing a test for dependence based on the entropy rate rather
than entropy, we are able to test for dependence among a group of
random variables, which is an improvement over pairwise tests proposed
in the literature to date. By using an asymptotically optimal compression
algorithm we avoid having to estimate joint densities to estimate
the entropy rate. Since general results for rates of convergence are
not yet available for the compression algorithm we use, we show how
a shuffling procedure can be used to nevertheless provide confidence
intervals for our entropy rate estimates. We discuss how our methodology
can be used in the important task of model selection, in particular
to identify appropriate lags for intertemporal models. We provide
strong evidence of the performance of this testing procedure under
full independence and full dependence. In particular, we see how this
test greatly outperforms procedures designed to detect only linear
dependencies, e.g. analysis of the auto-correlation function.

In future research, we would like to refine this test, study the estimator's
rate of convergence, and provide more evidence of its performance.
The potential applications of this estimator are very wide since one
only requires that the objects under investigation can be stored in
a computer and have a stable underlying stochastic structure.

\bibliographystyle{econometrica}
\bibliography{2013SV}

\section*{}
\end{document}